\documentclass[twocolumn,superscriptaddress,prb,showkeys,floatfix]{revtex4-2}
\usepackage{graphicx} 
\usepackage{amssymb}
\usepackage{amsmath}
\usepackage{xspace}

\begin{document}

\title{Multisubband Plasmons in InAs/GaSb Broken Gap Quantum Well}  

\author{Wojciech Julian Pasek}
\email{wojciech.pasek@um6p.ma}
\affiliation{School of Applied and Engineering Physics, University Mohammed VI Polytechnic, Ben Guerir, 43150, Morocco}

\author{Soufiane Hajji}
\email{soufiane.hajji@umontpellier.fr}
\affiliation{Institute of Electronics and Systems, University of Montpellier, CNRS, Montpellier 34095, France}

\author{Oussama Tata}
\affiliation{School of Applied and Engineering Physics, University Mohammed VI Polytechnic, Ben Guerir, 43150, Morocco}

\author{Laurent Cerutti}
\affiliation{Institute of Electronics and Systems, University of Montpellier, CNRS, Montpellier 34095, France}

\author{\mbox{Fernando Gonzalez-posada Flores}}
\affiliation{Institute of Electronics and Systems, University of Montpellier, CNRS, Montpellier 34095, France}

\author{\mbox{Simon Hurand}}
\affiliation{Institut Prime, Université de Poitiers 11 Boulevard Marie et Pierre Curie, Chasseneuil du Poitou 86360, France}

\author{\mbox{Abdelouahed El Fatimy}} 
\affiliation{School of Applied and Engineering Physics, University Mohammed VI Polytechnic, Ben Guerir, 43150, Morocco}

\author{Thierry Taliercio}
\email{thierry.taliercio@umontpellier.fr}
\affiliation{Institute of Electronics and Systems, University of Montpellier, CNRS, Montpellier 34095, France}

\begin{abstract}
Multi-subband plasmon (MSP) modes in heavily doped InAs/GaSb broken-gap quantum wells grown via molecular beam epitaxy (MBE) are investigated. An $8$-band $\vec{k} \cdot \vec{p}$ semiclassical model accurately predicts ellipsometric spectra, reflecting strong subband hybridization and non-parabolicity. In contrast, single-band plasmon models show qualitative discrepancies with experiment, even with adjusted effective masses. These findings highlight the potential of broken-gap wells for quantum technologies leveraging interband coupling and wavefunction hybridization.
\end{abstract}

\maketitle  

\newcommand{\kdp}{\vec{k}\cdot\vec{p}}
\newcommand{\dz}{\text{d}z}
\newcommand{\ket}[1]{\left|#1\right>}
\newcommand{\mel}[3]{\left\langle #1 \mid #2 \mid #3 \right\rangle}
\newcommand{\rcm}{~cm$^{-1}$\xspace}
\newcommand{\ekdp}{$8\times8~\vec{k}\cdot\vec{p}$ model\xspace}

\section{Introduction}
In recent years, the study of intersubband transitions (ISTs) and intersubband plasmons (ISP) in quantum wells (QW) has gained increasing interest due to their strong potential for applications in optoelectronics, particularly in quantum cascade lasers, infrared photodetectors, and electro-optical modulators \cite{choi1997physics,li2003interplay,cao2003interband}.
In a heavily doped QW, because several subbands can be filled by electrons, several ISPs can occur simultaneously. Contrary to predictions, experimental observations reveal that the optical response of such a system does not appear as a collection of distinct absorption peaks corresponding to all possible ISP. Instead, it manifests as a single absorption peak that combines the strength of all individual ISPs. The single absorption peak corresponds to a plasmonic mode known as the multisubband plasmon (MSP).

MSPs have been the subject of extensive study in contemporary research \cite{montesbajo2023perspectives}. In the literature, two theoretical models have been proposed to explain their formation: one, based on quantum mechanics \cite{todorov2012polaritonic,todorov2012intersubband,todorov2014dipolar,pegolotti2014quantum,askenazi2014ultra-strong,todorov2015dipolar,vasanelli2020semiconductor}, and the other, on a semi-classical approach \cite{alpeggiani2014semiclassical,montesbajo2018multisubband,pasek2022multisubband}. The quantum model expresses the light–matter interaction in the second quantization, starting from the electric displacement field and the polarization density (i.e. electric dipole gauge). A Bogoliubov diagonalization procedure is applied to capture the depolarization shift, which is an effect of electrons interacting among themselves in the scope of a single transition. Second Bogoliubov diagonalization follows yielding the MSP as the effect of coupling between different transition modes. The light–matter coupling component of the Hamiltonian is ultimately expressed in terms of effective plasma frequencies and overlap factors associated with the MSP modes. However, the semi-classical approach developed by Alpeggiani and Andreani \cite{alpeggiani2014semiclassical}, is based on quantum electrodynamics. The dielectric function was determined using Green's function and the transfer matrix formalism. Both models have been successfully used to calculate the dielectric function for certain systems, including the non-parabolicity of the bands. These calculations enable to understand the experimentally observed phenomena, such as the radiative decay rate of multi-subband plasmons \cite{choi1997physics} and the superradiant emission of a collection of particles \cite{dicke1954coherence}. Despite their success, both models fail to accurately describe a system characterized by strong non-parabolicity and mixing between valence and conduction quantum states, such as in the case of InAs/GaSb QW.

Sb-based heterostructures such as InAs/GaSb or InAs/AlSb and their alloys are of interest for two main reasons. Firstly, they are almost lattice-matched to GaSb or InAs substrates. This allows thick heterostructures to be grown with limited strain relaxation. Secondly, they offer all kinds of band alignments, from type-I, where electrons and holes are localised in the same material, to type-III, where electrons and holes are localised in neighbouring materials (well and barrier) and the conduction and valence bands of both materials are resonant. Sb-based heterostructures, which allow the fabrication of quantum cascade lasers \cite{loghmari2019continuous}, interband cascade lasers \cite{díaz-thomas2025single} and type-II superlattice photodetectors \cite{bouschet2023electro-optical}, are best suited for photonic applications in the mid-IR. In addition, they open up interesting areas of research such as topology \cite{avogadri2022large}.

In this work, a generalization of the semi-classical model is proposed by including the coupling between valence and conduction quantum states and non-parabolicity. This generalization is based on the $8\times8~\vec{k}\cdot\vec{p}$ formalism. The resulting model can be applied to all types of QW without limitations.

This paper is organized as follows. In Section~\ref{sec:samples}, the studied samples are described; these are type-III broken-gap InAs/GaSb QW nanostructures. In Section~\ref{sec:spectroscopic_ellipsometry}, the results of spectroscopic ellipsometry measurements are presented and discussed. In Section~\ref{sec:single-particle_schrödinger-poisson_equation}, the single-particle Hamiltonians considered in this work are introduced: the single-band one and the $8\times8~\vec{k}\cdot\vec{p}$ one. The eigensolutions of these Hamiltonians are subsequently used as the foundation for the plasmon models developed in the following section. In Section~\ref{sec:semiclassical_plasmon_models}, a semiclassical theory of the intersubband response of quantum wells is presented. The parabolic model is introduced first, which is known to efficiently describe simple type-I quantum well systems. This is followed by the hybrid model, in which a position-dependent effective mass is incorporated while remaining within the single-band framework. Finally, a full \ekdp is described, in which intersubband response is treated by including conduction and valence band mixing, as well as non-parabolicity.
The paper concludes with Sections~\ref{sec:conclusion}, \ref{sec:discussion}, and \ref{sec:Outlook_and_perspectives}, which present the conclusions, discussion, and outlook for future developments, respectively.

\section{Samples}\label{sec:samples}
The QW structures were grown by solid source molecular beam epitaxy (MBE) in Riber 412 MBE system, equipped with valved Arsenic (As) and antimony (Sb) cracker cells.

\begin{figure*}[hbt!]
\centering
\includegraphics[width=0.8\linewidth]{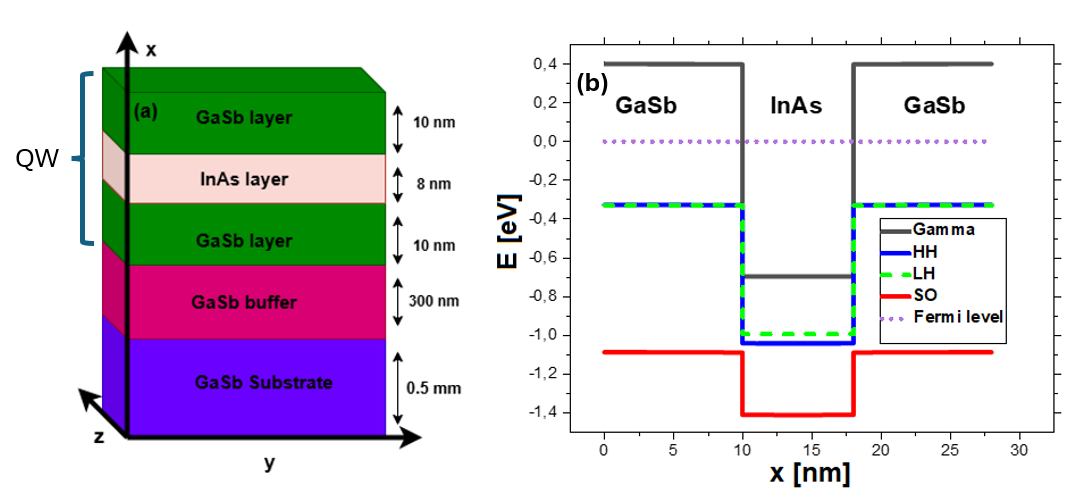}
\caption{(a) Growth scheme of InAs/GaSb QW. (b) Energy band profile of the  QW. Gamma, HH, LH, and SO  refer respectively to the conduction band, the heavy hole band, the light hole band, and the spin-orbit split-off band. Fermi level is taken as the reference energy level.} 
\label{fig:gr}
\end{figure*}
Fig.~\ref{fig:gr}(a) presents the growth scheme. A $0.5$ mm layer of Te-doped GaSb $( ~1\times 10^{18}$ $\text{cm}^{-3})$
 was used as a \mbox{substrate}. To bury lattice mismatch induced defects, a $300$~nm GaSb buffer layer was grown on the substrate. finally, the InAs/GaSb QW was grown. Two samples were fabricated:
\begin{itemize}
    \item \textbf{Sample (A)}:\\
    $L_{\text{InAs}} = 8$ nm and $n_e = 1.22 \times 10^{19}$ cm$^{-3}$
    \item \textbf{Sample (B)}:\\
    $L_{\text{InAs}} = 18$ nm and $n_e = 1.16 \times 10^{19}$ cm$^{-3}$
\end{itemize}
The doping level is adjusted by monitoring the silicon cell temperature and using oblique incidence reflectance experiments under polarized light \cite{taliercio2014brewster}. Fig.~\ref{fig:gr}(b) represents the potential confinement profile of the electrons in the InAs layer. This type of energy band alignment leads to a conduction–valence state mixing effect Ref.~\cite{zakharova2001hybridization}.

\section{Spectroscopic ellipsometry}\label{sec:spectroscopic_ellipsometry}
\begin{figure*}[hbt!]
\centering
\includegraphics[width=0.8\linewidth]{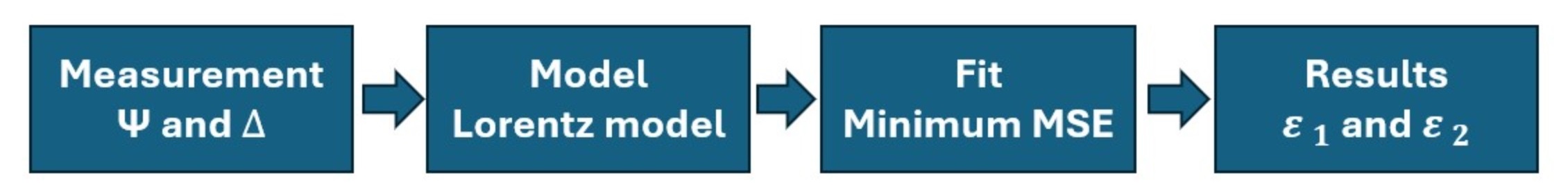}
\caption{Procedure followed to extract the optical constants.}
\label{Algo}
\end{figure*}

\begin{figure*}[hbt!]
\centering
\includegraphics[width=0.8\linewidth]{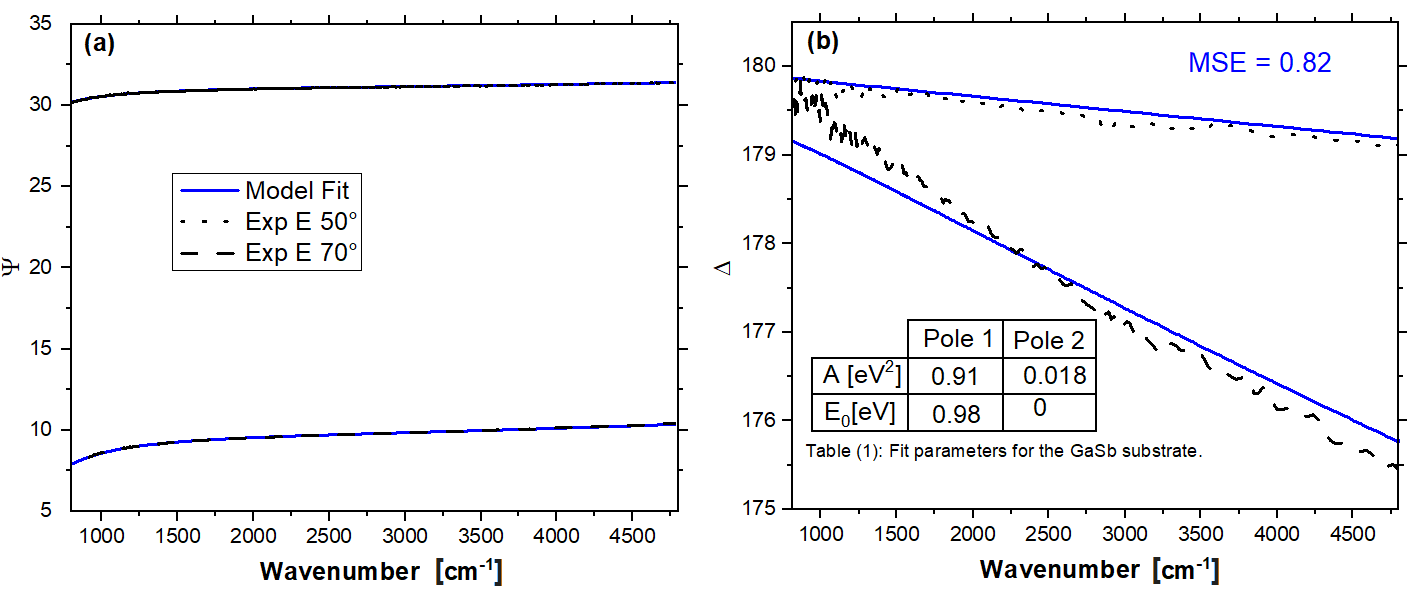}
\caption{The modeled (solid line) and measured (dashed/dotted lines) $\psi$ [in (a)] and $\Delta$ [in (b)] for the GaSb substrate. The table inset in (b) presents the fit parameters for the GaSb substrate.}
\label{sub_fit}
\end{figure*}

\begin{figure*}[hbt!]
\centering
\includegraphics[width=0.8\linewidth]{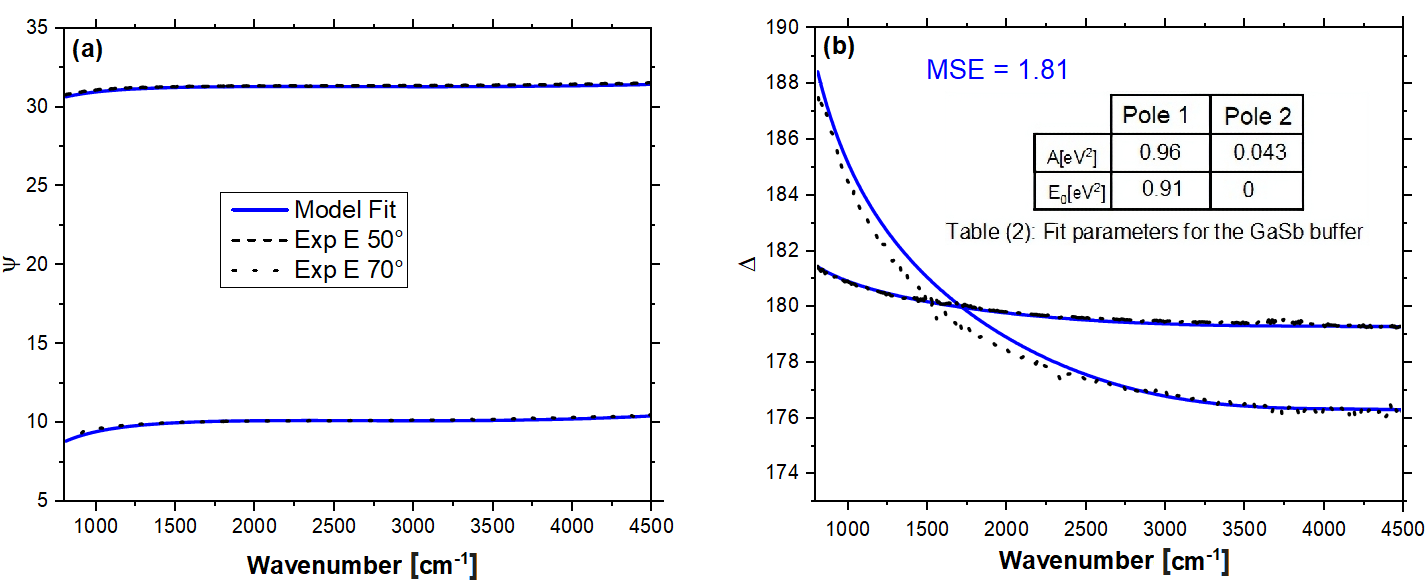}
\caption{The modeled (solid line) and measured (dashed/dotted lines) $\psi$ [in (a)] and $\Delta$ [in (b)] for the GaSb buffer deposited on a GaSb substrate. The table inset in (b) presents the fit parameters for the GaSb buffer.}
\label{buffer_fit}
\end{figure*}

\begin{figure*}[hbt!]
\centering
\includegraphics[width=0.8\linewidth]{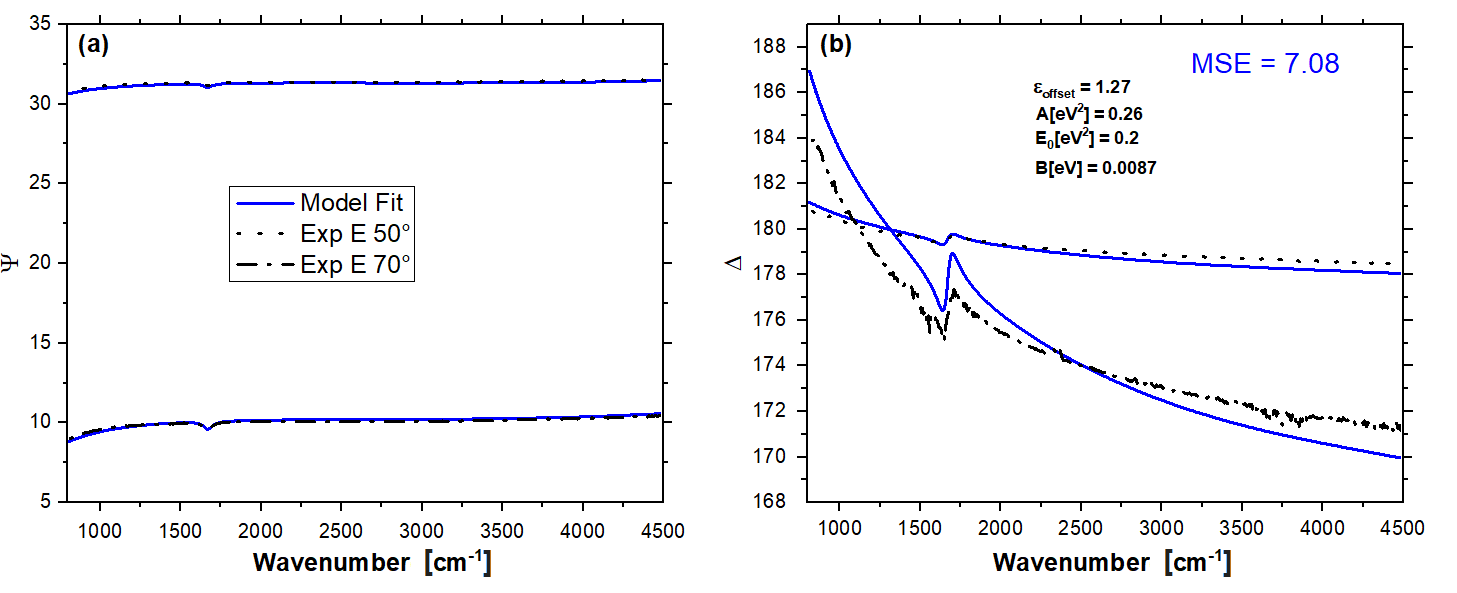}
\caption{The modeled (solid line) and measured (dashed/dotted lines) $\psi$ [in (a)] and $\Delta$ [in (b)] for the InAs/GaSb QW deposited on the GaSb buffer and the GaSb substrate. Parameters of the single Lorentz oscillator of Eq.~(\ref{lor}) are given in (b).}
\label{fit_QW}
\end{figure*}

\begin{figure*}[hbt!]
    \centering
    \includegraphics[width=0.497\linewidth]{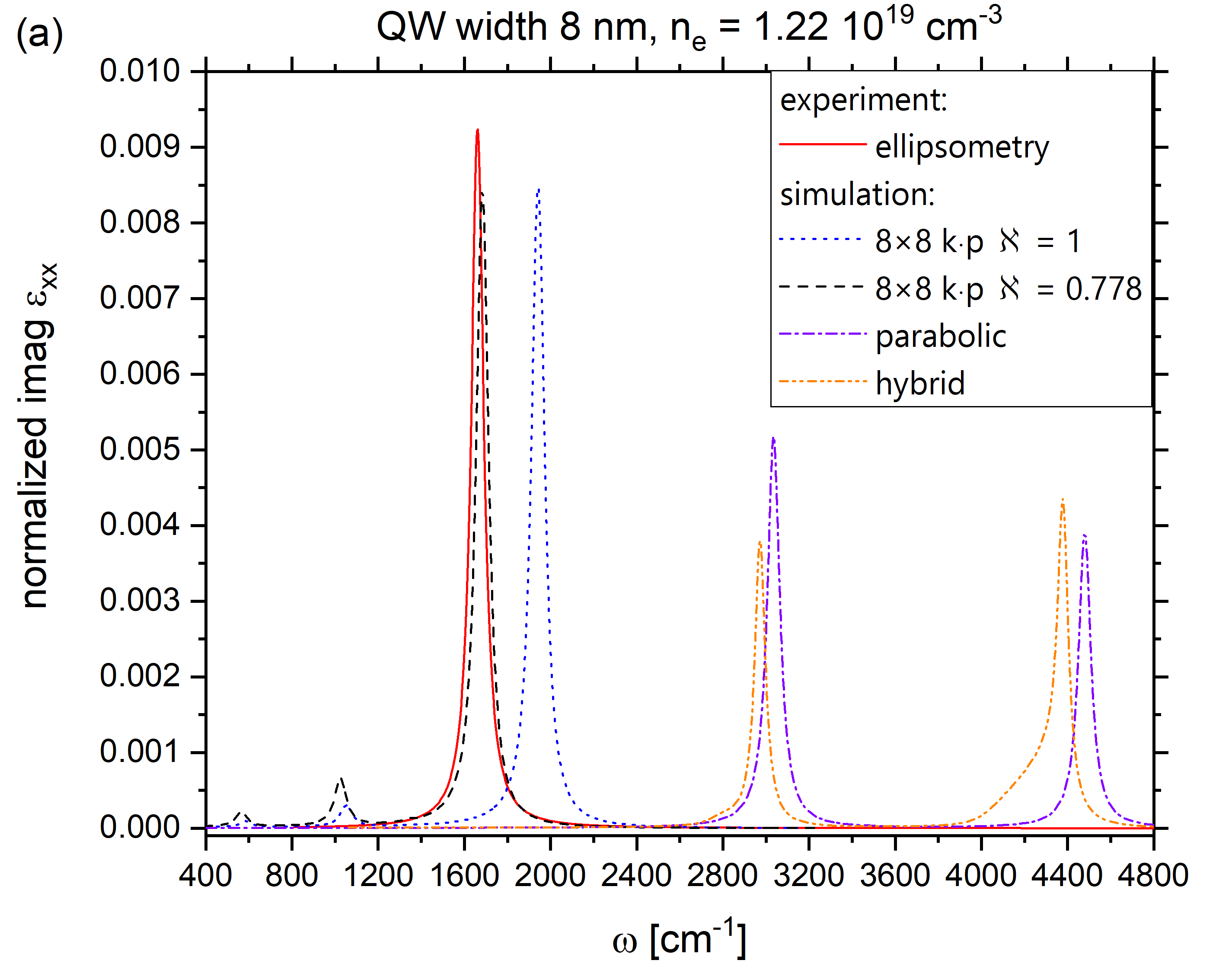}
    \includegraphics[width=0.497\linewidth]{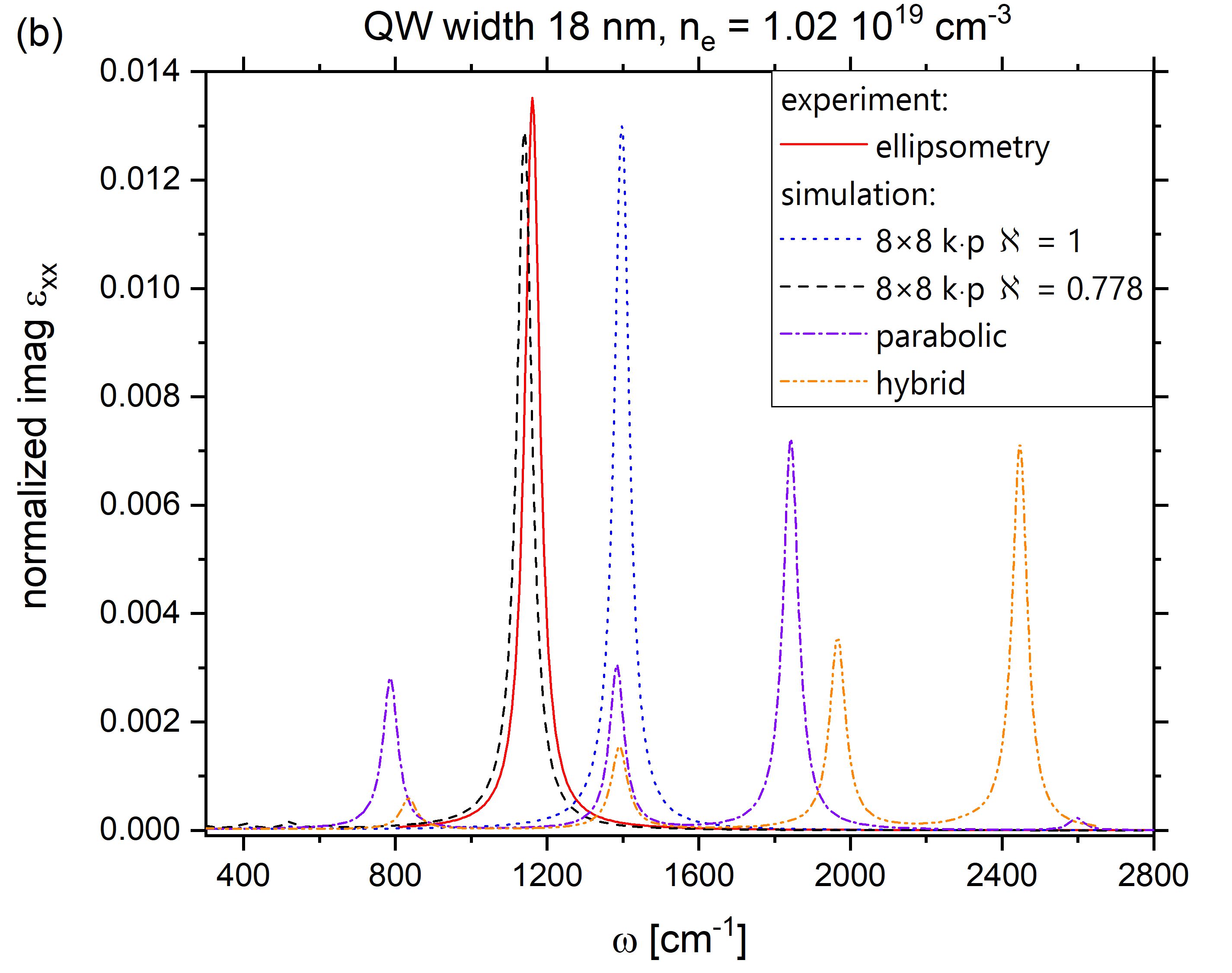}
    \caption{The peaks of the imaginary part of the effective dielectric function $\epsilon_{xx}$ identified experimentally by the ellipsometry experiment (red solid line) and the corresponding spectra obtained from all the semiclassical models (other line styles). Results shown in (a) are for the $8$~nm QW system, the ones shown in (b) correspond to the $18$~nm QW one. Spectra were normalized to facilitate direct comparison between the shapes.}
    \label{fig:model_comparison}
\end{figure*}
The optical properties of the QW were characterized with a variable-angle spectroscopic ellipsometer in the infrared range. Ellipsometry is an optical technique that involves illuminating a sample with polarized light and measuring the change in polarization state of the reflected light by the sample \cite{woollam1990fundamentals}. The polarization change is quantified by two quantities: the amplitude ratio $\Psi$ and the phase difference $\Delta$, which are related to the Fresnel reflection coefficients $R_s$ and $R_p$, corresponding to s-polarized and p-polarized light, respectively, through the following relation:

\begin{equation}
  \frac{R_p}{R_s}=\tan(\Psi)\exp{(i\Delta)}
\end{equation}
In the case of a heterostructure, $\Psi$  and $\Delta$ cannot be converted directly into the optical constants. In this context, the optical constants are determined by the process outlined in Fig.~\ref{Algo}. The experimental data $\Psi$  and $\Delta$ are obtained from $1.7$ to $40$~$\mu\text{m}$ using an IR-VASE instrument from J.A. Woollam. The optical model was developed using the WVASE software \cite{woollam2012guide}. The optimal model is the one that minimizes the mean square error (MSE) between the $\Delta$ and $\Psi$  values calculated by the optical model and those obtained experimentally. 
In order to extract the optical constants of the QW, the substrate was first characterized to determine its optical properties. Then, the GaSb buffer layer was studied. The parameters obtained from the substrate and the buffer layer were then used to model the entire structure (substrate+buffer+QW), which finally allows to deduce the optical constants of the QW layer, which refers here to a single optical layer ($28$ or $38$~nm thick), comprising the three sublayers (GaSb ($10$~nm) + InAs ($8$ or $18$~nm) + GaSb ($10$~nm)).\\
Before taking the measurements, the back side of the GaSb substrate of all samples was carefully depolished using $5$-micron alumina particles. 
This step was implemented to eliminate unwanted light reflections from the back side of the GaSb substrate. Such reflections can disrupt the polarization of the measured light, thereby compromising the accuracy of the obtained results. The experimental data for all samples were collected at two incidence angles ($50^{o}$ and $70^{o}$) under optimized conditions to improve the quality of the results. The number of scans was increased to $500$ to improve the signal-to-noise ratio, and the resolution was set to $8$~$cm^{-1}$ to resolve fine spectral features. To construct the  optical model, the general oscillator layer implemented in the WVASE software was used.\\
For the GaSb substrate the real part of the dielectric function $\epsilon_{1}$ is modeled using two poles. Each pole corresponds to a Lorentz oscillator without broadening, which is formally equivalent to a Sellmeier oscillator. These oscillators simulate the dispersion in $\epsilon_{1}$ created by absorption that occurs outside the measured spectral range~\cite{woollam2012guide}.
\begin{equation}
  \epsilon_{1}=\epsilon_\text{offset} +\frac{A_{1}}{E_{1}^{2}-E^{2}}+\frac{A_{2}}{E_{2}^{2}-E^{2}}
\end{equation}
Here $\epsilon_\text{offset}$ is a purely real constant and is fixed at 13.13. The pairs ($A_{1}$, $A_{2}$) and  ($E_{1}$, $E_{2}$) correspond to the amplitudes and energy centers of the poles on the visible side and on the infrared side, respectively. The numerical values of these parameters are summarized in Fig.~\ref{sub_fit}.
Within the spectral range studied ($200-3000$\rcm), the substrate is considered transparent, implying that the imaginary part of its dielectric function $\epsilon_{2}$ is negligible. Surface oxidation is modeled by a roughness layer \cite{ronnow1995surface}. Fig.~\ref{sub_fit} shows the measurement results of $\psi$ and $\Delta$, as well as the corresponding fitting results as the inset in (b). MSE of 0.82 indicates an excellent fit to the experimental data.  For the GaSb buffer layer is modeled using a similar approach. The fit parameters are given in  Fig.~\ref{buffer_fit}(b). The integration constant $\epsilon_\text{offset}$  is set to $13.21$. Fig.~\ref{buffer_fit} also represents the measured $\psi$ and $\Delta$ data along with their corresponding fit. MSE = 1.81 indicates a good fit to the experimental data. The GaSb buffer layer exhibits poles different from those of the GaSb substrate, indicating a difference in doping levels. This discrepancy between the two GaSb layers highlights the importance of characterizing each layer individually, as done in this study.
  The results of the modeling of the substrate and the buffer will serve as tools for extracting the dielectric function of the InAs/GaSb QW.\\ The dielectric function of the QW layer is modeled as an isotropic layer using a single Lorentz oscillator. 
\begin{equation}
  \epsilon= \epsilon_\text{offset}+\frac{iA}{E_{0}^{2}-E^{2}-iBE}
  \label{lor}
\end{equation}
For sample (A), the corresponding values of amplitude (A), energy center ($E_0$), and broadening (B) are presented in Fig.~\ref{fit_QW}. The same figure represents the measured $\Psi$ and $\Delta$  data along with their corresponding fit. MSE = 7.08, which is considered a fairly good fit. Note that sample (B) was modeled using the same method. Fig.~\ref{fig:model_comparison} represents the imaginary part of the dielectric function derived from model Eq.~(\ref{lor}). 
The analysis of the ellipsometry data highlights the formation of the MSP in InAs/GaSb QW. For sample (A), a single MSP mode is predicted at $1661$ \rcm, while for sample (B), a single MSP mode appears at $1161$ \rcm. The  MSP in sample (B) is shifted compared to sample (A) because the MSP depends on the doping density and the thickness of the QW. 

\section{Single-particle Schrödinger-Poisson equation}\label{sec:single-particle_schrödinger-poisson_equation}

The simulation of the plasmon modes in QW heterostructure is a two-step process. It begins with obtaining the dispersion relations $E_{n,\vec{k}}$ and corresponding wavefunctions $\Psi_n(x,\vec{k})$ of a single electron. At this stage, the energies of individual transitions between bands $n$ and $m$ denoted by $E_{n,n+1}(\vec{k})$ and the associated frequencies $\omega_{n,m,\vec{k}}$ are obtained as $E_{n,m}(\vec{k}) = E_{m,\vec{k}} - E_{n,\vec{k}} = \hbar \omega_{n,m,\vec{k}}$. In the mentioned relations, $x$ marks the direction of growth, while $\vec{k} = \left( k_y, k_z \right)$ is the in-plane wavevector. For simplicity, the initial and final band indices $n$ and $m$ can be reformed into a single index set $\alpha \equiv \lbrace n, m \rbrace$. Diagonalization of the $8$-band $\kdp$ single-particle Hamiltonian is performed using \textit{nextnano++} software \cite{birner2007nextnano}.

These single-particle results form the starting point for the second stage of the computation, that is the semi-classical plasmon model. The output of the latter is the effective susceptibility $\chi(\omega)$ of the nanosystem, which in turn straightforwardly translates to the effective dielectric constant $\epsilon(\omega)$ with the help of a simple effective medium approach. The imaginary part of this function, $\Im\left(\epsilon(\omega)\right)$, is a quantity which, on one hand, characterizes the absorption properties of the system and, on the other hand, is directly obtainable by ellipsometry spectroscopy.

\subsection{single-band single-particle Hamiltonian}\label{sec:single-band_single_particle}

\begin{figure*}[hbt!]
    \centering
    \includegraphics[width=0.32\linewidth]{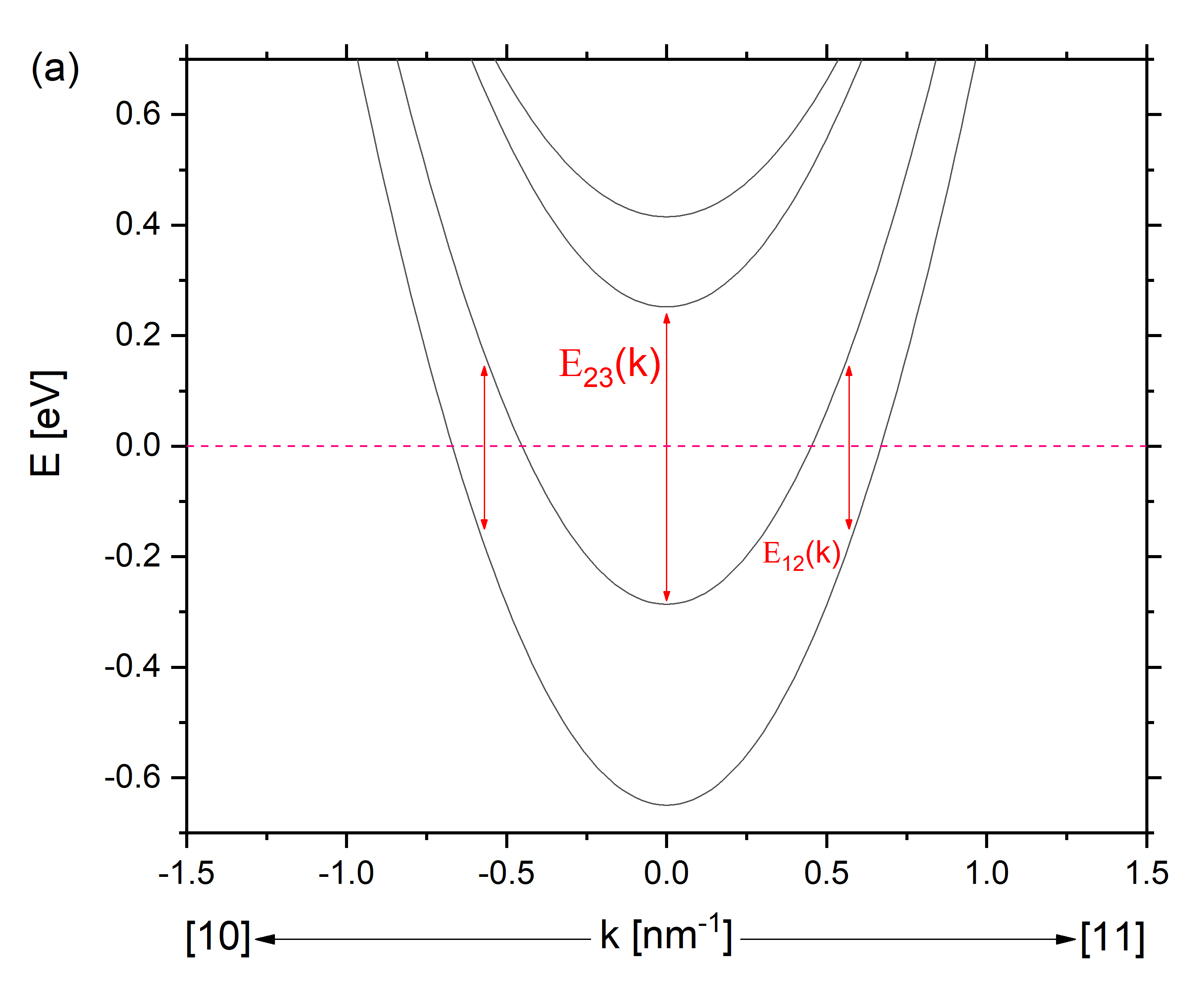}
    \includegraphics[width=0.32\linewidth]{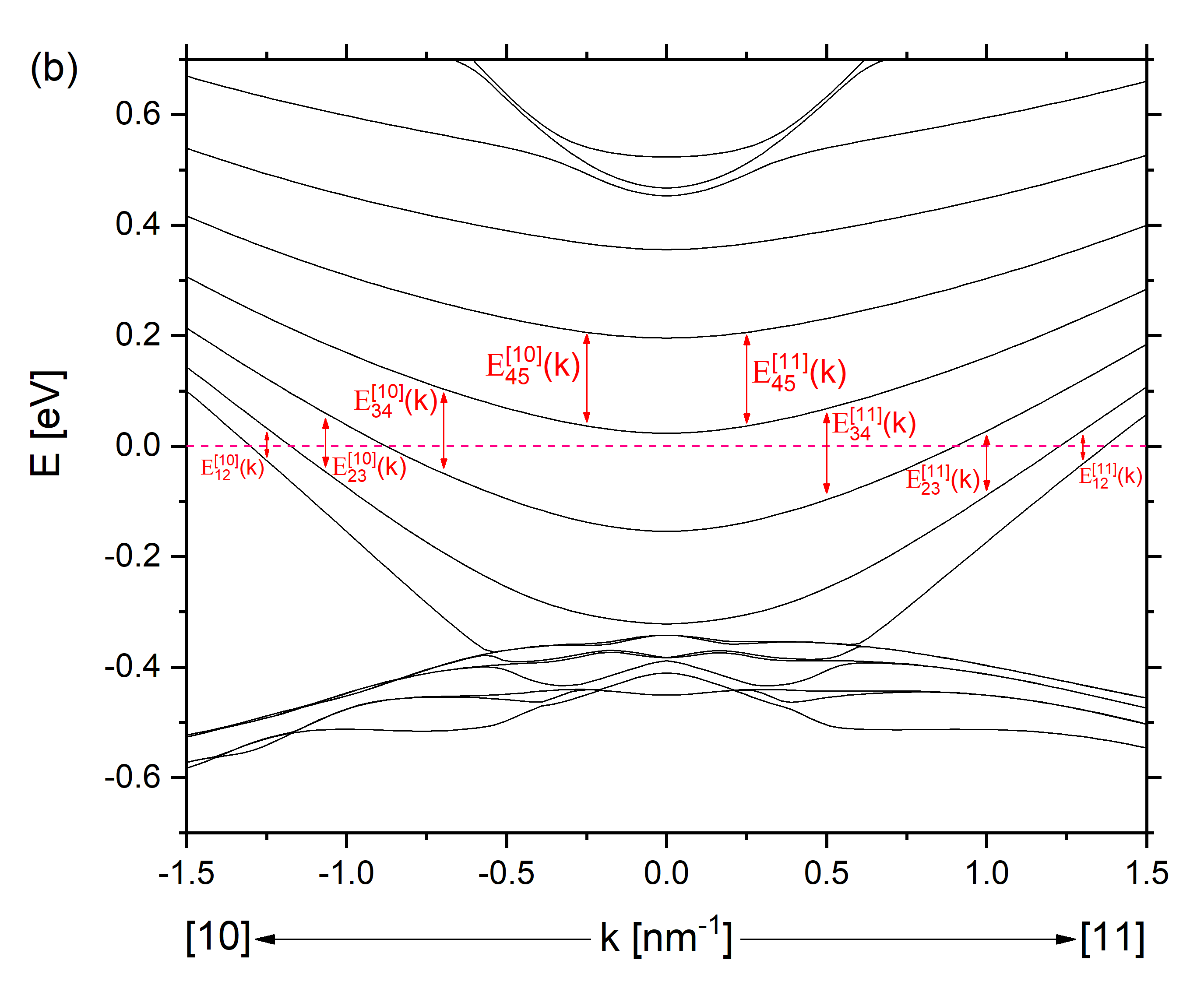}
    \includegraphics[width=0.32\linewidth]{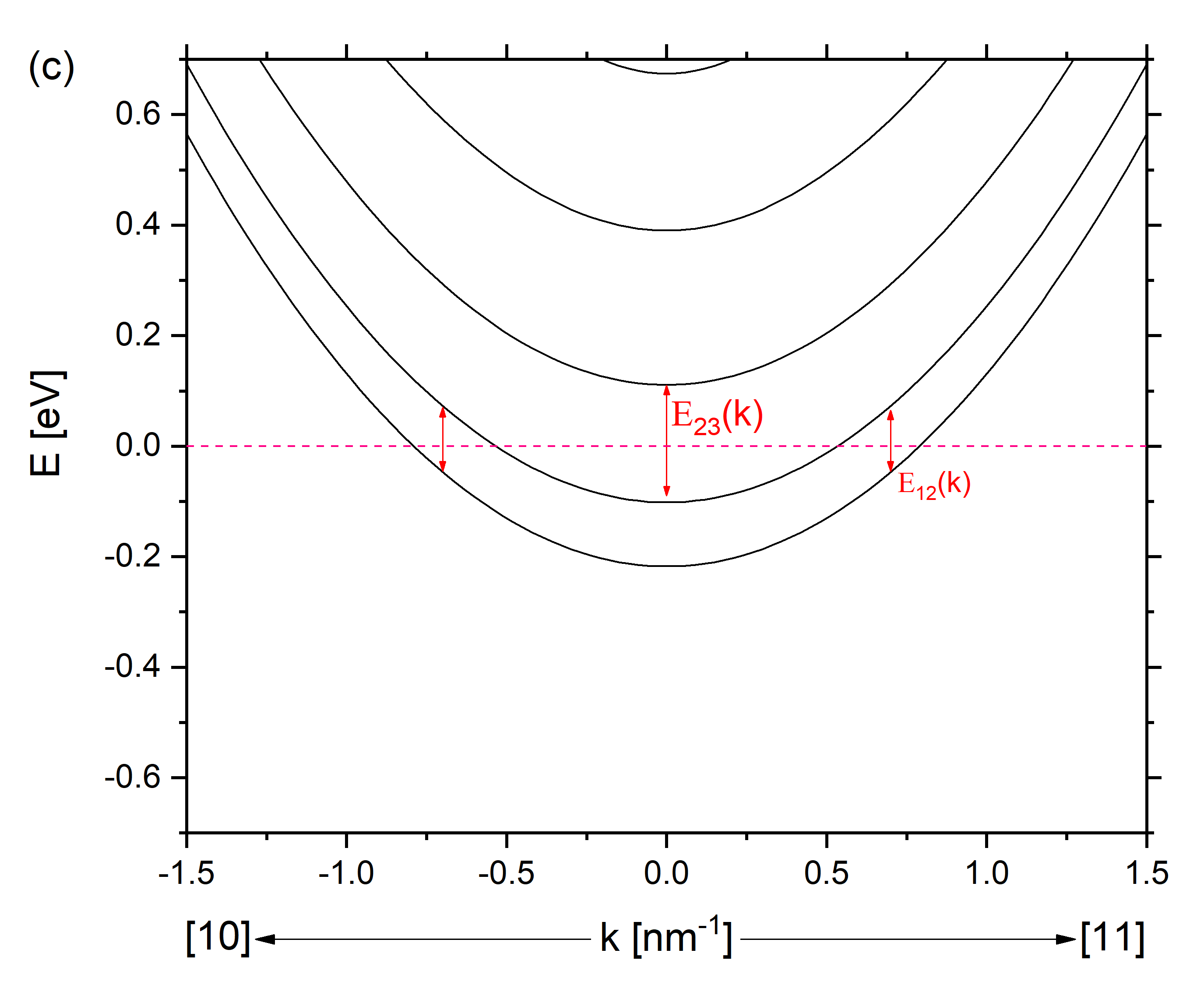}
    \caption{Comparison of the single-particle dispersion relations for the $8$~nm QW system obtained with different models: (a)~the~single-band model with the original bulk effective mass, (b)~the \ekdp, (c)~the single-band model with an arbitrary effective mass $\tilde{m}_\text{InAs}^* = 0.11\,m_0$ fitted to the experimental results. Note that the range of energy shown is the same in (a), (b) and (c), but the scale of energy separations of the neighboring subbands in the relevant part of the spectrum (vertical red arrows) seen in (a) differs significantly from these in (b)/(c). The dashed horizontal line at zero energy corresponds to the Fermi level.}
    \label{fig:8nm_dispersion_comparison}
\end{figure*}

In this work, two models for the single electron system will be taken into account, the simple single-band effective mass Hamiltonian and the \ekdp one. In fact, the twofold spin degenerate single-band Schrödinger equation for the electrons in a heterostructure is a special case of the \ekdp, applicable in the case when the coupling between the conduction and valence bands is ignored. It has the form of:
\begin{equation}
    \left[-\frac{\hbar^2}{2} \nabla \cdot M(x) \nabla -q\phi(x)+ V(x) \right] \psi_n(x) = E_n \psi_n(x),
\end{equation}
where the potential part $V(x)$ comes from the relative offset in energy, between the well and the barrier materials, of the conduction bands at the $\Gamma$ point. $\phi(x)$ is the electrostatic potential given by Poisson equation. The inverse effective mass tensor is, in general, represented by a $3{\times}3$ matrix, with elements of the form $M_{r,s} = 1/m_{r,s}$, where $r, s \in \lbrace x, y, z \rbrace$. In the case of the system in this work, a simple isotropic effective mass is used, due to the spherical symmetry of the $S$ orbital of the conduction bands. 
However, this isotropic mass is not homogeneous, $M \equiv M(x)$, as it has different values in the well and in the barrier, and thus depends on the $x$ coordinate along the growth axis. This results in non-parabolicity in the form of dependence of the band curvature on the magnitude of the in-plane vector $\vec{k}$ and the resulting dependence of the energy separation of neighboring bands $E_{\alpha}(k)$. In the simplest parabolic model of the MSP formation, this dependence is neglected and the system is assumed to be fully characterized by its wavefunctions and energies at the $\Gamma$ point $(\vec{k}=0)$. Conversely, they are taken into account in the hybrid model.
Given the solution of the Schrödinger equation, the electrostatic potential $\phi(x)$ can be obtained as
\begin{equation}
    \nabla\cdot\left[ \epsilon_0 \epsilon_r(x) \nabla \phi(x) \right] = -\rho(x),
\end{equation}
from the corresponding charge density distribution
\begin{equation}
    \rho(x) = e \left[ -n(x) + p(x) + N_D(x) - N_A(x) \right],
\end{equation}
where $n$ and $p$ are the electron and hole densities, and $N_D$ and $N_A$ are the ionized donor and acceptor concentrations, respectively. Note that, in the case of the single-band model taking into account only the conduction band, the $p(x)$ term is identically equal to zero, the $N_D(x)$ corresponds to intentional bulk Si doping of the well, and the $N_A(x)$ term to the residual unintentional $p$-type doping in the barrier.
However, since the electrostatic potential $\phi(x)$ and the charge density $\rho(x)$ are mutually dependent, the system should be solved self-consistently by a relaxation iterative procedure. The details of the implementation of the computation are given in  Ref.~\cite{birner2011modeling}. The material parameters for InAs and GaSb were taken from Ref.~\cite{vurgaftman2001band} and are detailed in Appendix~\ref{sec:material_parameters}.

\begin{figure*}[hbt!]
    \centering
    \includegraphics[width=0.32\linewidth]{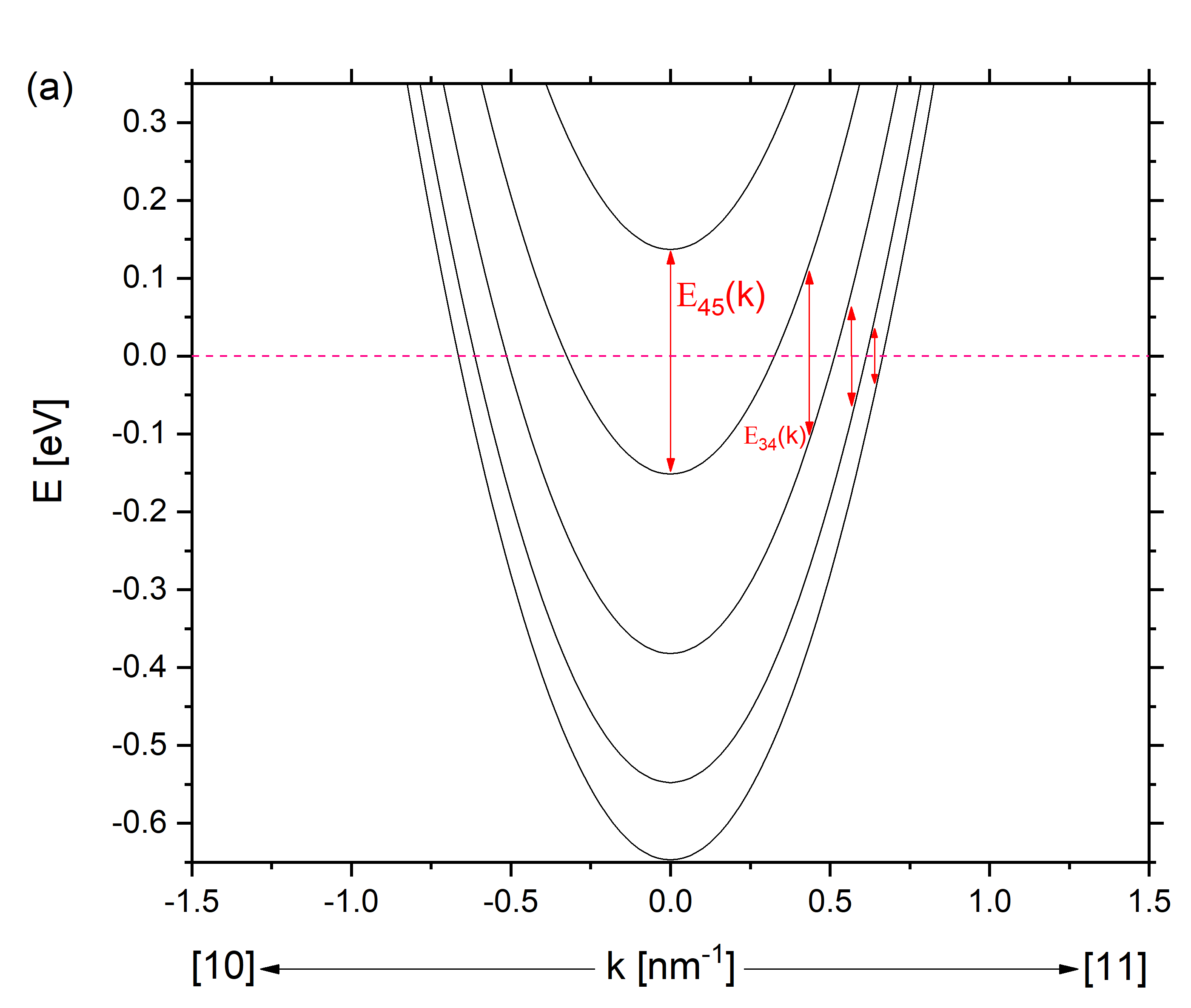}
    \includegraphics[width=0.32\linewidth]{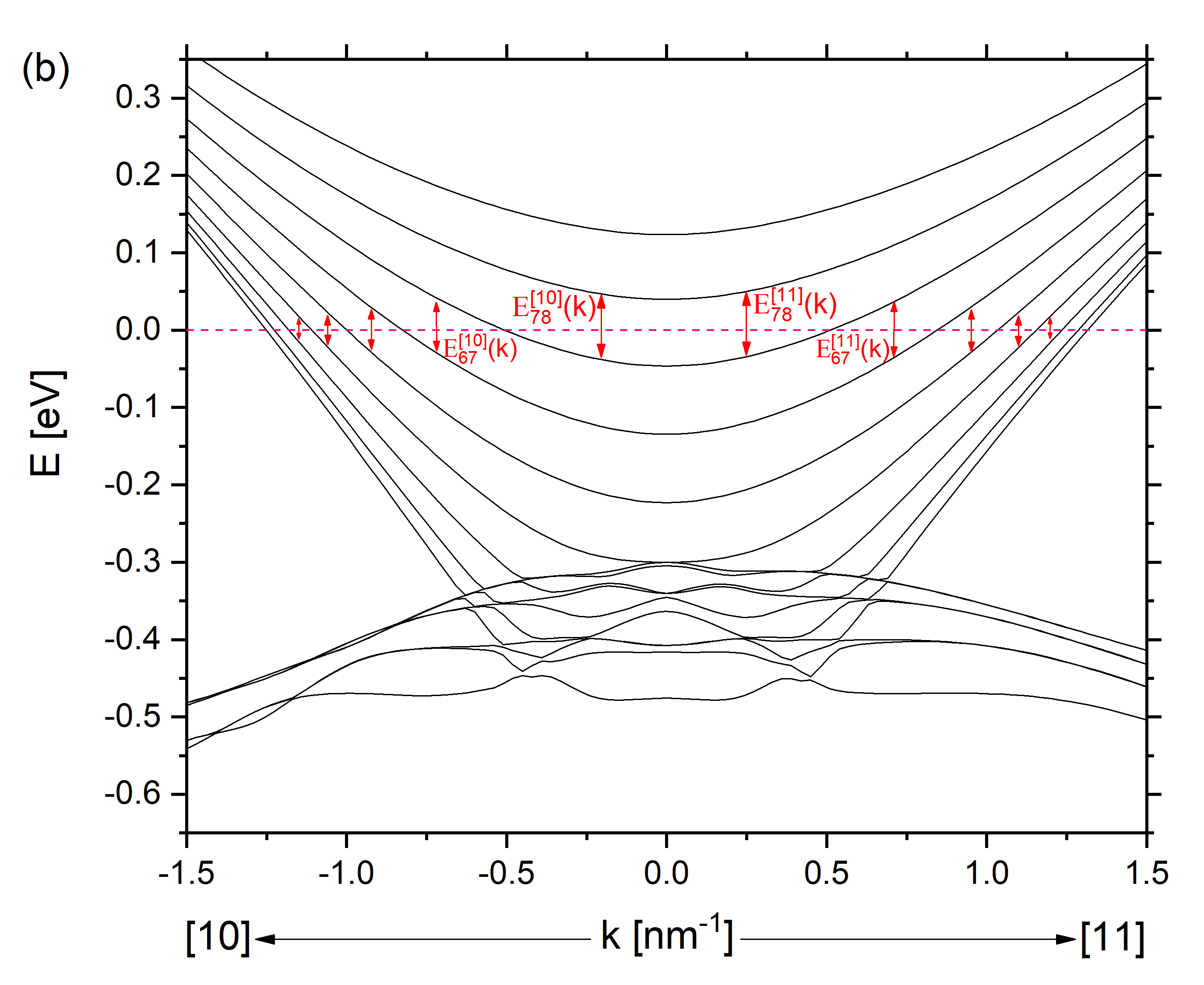}
    \includegraphics[width=0.32\linewidth]{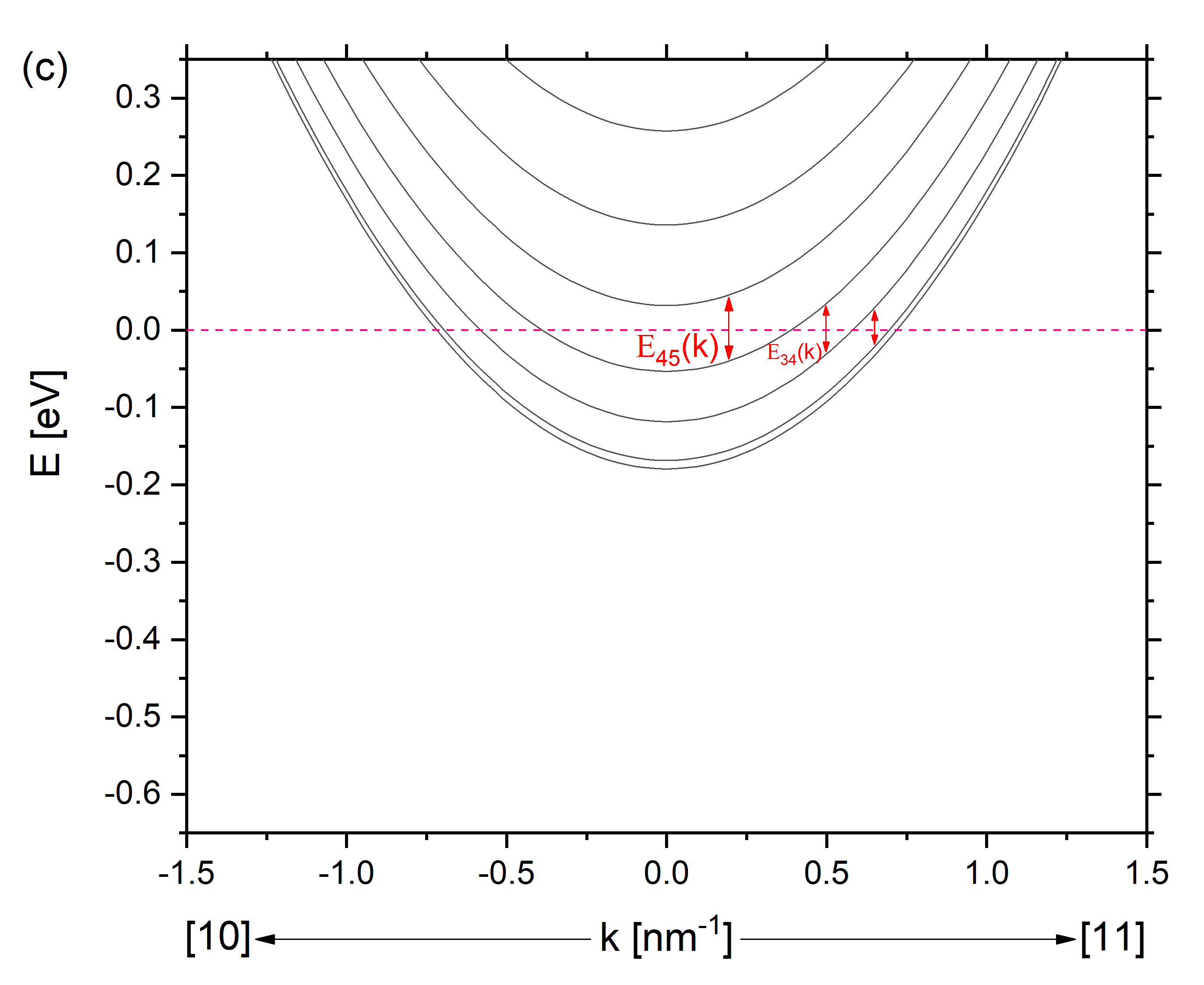}
    \caption{Comparison of the single-particle dispersion relations for the $18$~nm QW system obtained with different models: (a)~the~single-band model with the original bulk effective mass, (b)~the \ekdp, (c)~the single-band model with an arbitrary effective mass $\tilde{m}_\text{InAs}^* = 0.11\,m_0$ fitted to the experimental results. Note that the range of energy shown is the same in (a), (b) and (c), but the scale of energy separations of the neighboring subbands in the relevant part of the spectrum (vertical red arrows) seen in (a) differs significantly from these in (b)/(c). The dashed horizontal line at zero energy corresponds to the Fermi level.}
    \label{fig:18nm_dispersion_comparison}
\end{figure*}

Figures \ref{fig:8nm_dispersion_comparison}(a) and \ref{fig:18nm_dispersion_comparison}(a) show the subband dispersion of $8$~nm QW and $18$~nm QW systems, respectively, with the Fermi level taken as the energy reference. Note that in this case the dispersion along $\vec{k}$ directions of $[10]$ and $[11]$ is identical, as $E_{n,\vec{k}} \equiv E_{n,k}$. It can be noticed that, even in the case of the simple single-band Hamiltonian, the energy separations between the neighboring subbands depend strongly on $k$, being maximal for each transition $\alpha$ at the $\Gamma$ point and diminishing as $k = \left| \vec{k} \right|$ increases. In practice, the important energy separations are the ones corresponding to transitions between one filled subband lying below $E_F$ and the other empty subband lying above $E_F$, as shown by the vertical red arrows in Figs.~\ref{fig:8nm_dispersion_comparison}(a) and \ref{fig:18nm_dispersion_comparison}(a).

\subsection{$8$-band $\kdp$ single-particle Hamiltonian}\label{sec:k.p_single_particle}
\begin{widetext}
\small\begin{equation}
\hat{H}_{8\times8} = 
   \begin{bmatrix}
     \vspace{3pt}\\
     C & 0 & \mathcal{P}_{y,z,x} & \mathcal{P}_{z,x,y} & \mathcal{P}_{x,y,z} & 0 & 0 & 0\vspace{3pt}\\
     0 & C & 0 & 0 & 0 & \mathcal{P}_{y,z,x} & \mathcal{P}_{z,x,y} & \mathcal{P}_{x,y,z}\vspace{3pt}\\
      &  & \mathcal{M}_{x,y,z} & \mathcal{N}_{x,y} - \frac{i\Delta}{3} & \mathcal{N}_{x,z} & 0 & 0 & \frac{\Delta}{3}\vspace{3pt}\\
      &  &  & \mathcal{M}_{y,x,z} & \mathcal{N}_{y,z} & 0 & 0 & - \frac{i\Delta}{3}\vspace{3pt}\\
      &  &  &  & \mathcal{M}_{z,x,y} & -\frac{\Delta}{3} & \frac{i\Delta}{3} & 0\vspace{3pt}\\
      &  &  &  &  & \mathcal{M}_{x,y,z} & \mathcal{N}_{x,y} + \frac{i\Delta}{3} & \mathcal{N}_{x,z}\vspace{3pt}\\
      &  &  &  &  &  & \mathcal{M}_{y,x,z} & \mathcal{N}_{y,z}\vspace{3pt}\\
      &  &  &  &  &  &  & \mathcal{M}_{z,x,y}
   \end{bmatrix},
\label{eq:k.p_matrix}\end{equation}\normalsize
\end{widetext}
Equation (\ref{eq:k.p_matrix}) presents the full \ekdp Hamiltonian expressed in the basis $\ket{S\uparrow}$ $\ket{S\downarrow}$, $\ket{X\uparrow}$, $\ket{Y\uparrow}$, $\ket{Z\uparrow}$, $\ket{X\downarrow}$, $\ket{Y\downarrow}$, $\ket{Z\downarrow}$, as described in Ref.~\cite{birner2011modeling}, where the matrix elements are defined by
\small\begin{equation}
    C = E_C + \frac{\hbar}{2 m_0} (\hat{k}_x S \hat{k}_x + \hat{k}_y S \hat{k}_y + \hat{k}_z S \hat{k}_z),\nonumber
\end{equation}
\begin{equation}
    \mathcal{N}_{{r_1},{r_2}} = \hat{k}_{r_1} N^+ \hat{k}_{r_2} + \hat{k}_{r_2} N^- \hat{k}_{r_1},\nonumber
\end{equation}
\begin{equation}
    \mathcal{P}_{{r_1},{r_2},{r_3}} = \hat{k}_{r_1} B \hat{k}_{r_2} + i P \hat{k}_{r_3},\nonumber
\end{equation}
\begin{equation}
    \mathcal{M}_{{r_1},{r_2},{r_3}} = E_V + \frac{\hbar}{2 m_0}\hat{k}^2 + \hat{k}_{r_1} L \hat{k}_{r_1} + \hat{k}_{r_2} M \hat{k}_{r_2} + \hat{k}_{r_3} M \hat{k}_{r_3},
\end{equation}\normalsize
where $\lbrace r_1,r_2,r_3 \rbrace$ is a permutation of $\lbrace x,y,z \rbrace$ and the lower triangular part can be obtained from the hermicity of $\hat{H}_{8\times8}$. The diagonal conduction band material parameter can be obtained as
\begin{equation}
    S = 1 + \frac{2}{m_0} \sum_n \frac{|\mel{S}{p_x}{n}|^2}{E_C - E_n},   
\label{eq:S_renormalization}\end{equation}
where the sum in the second term is done over the remote bands. The Kane inter-band matrix element is $P = \frac{\hbar}{m_0} \mel{S}{\hat{p}}{X}$. Finally, $L, M, N^+, N^-$ are the Dresselhaus–Kip–Kittel material parameters for the valence band, see Ref.~\cite{dresselhaus1955cyclotron}.

This model allows for a proper description of the conduction-valence band mixing and of the creation of hybridized wavefunctions due to quantum size effect, which is especially relevant in type-III (broken gap) nanostructures. If the $x$ axis is the growth axis of the heterostructure, then $\hat{k}_x = -i\frac{\partial}{\partial x}$, while $\hat{k}_y = k_y$ and $\hat{k}_z = k_z$. Consequently, the model results in energy separations $\omega_{\alpha}(\vec{k})$ and $8$-component wavefunctions $\psi_{n,\vec{k}}(x)$, which are in general essentially dependent on the in-plane wavevector $\vec{k}$, both in the terms of its magnitude and direction.

Note that the mixing between the conduction and the valence bands is an effect of the off-diagonal term $\mathcal{P}_{{r_1},{r_2},{r_3}}$ in Eq.~(\ref{eq:k.p_matrix}). For the specific materials of the studied heterostructure $B=0$ and this term takes the explicit form of $\mathcal{P}_{{r_1},{r_2},{r_3}} \equiv \mathcal{P}_{r_3} = i P \hat{k}_{r_3}\nonumber$. In the case of bulk material, $\mathcal{P}_{r_3}$ vanishes at the $\Gamma$ point of the bulk Brillouin zone $\vec{k}_b=(k_x,k_y,k_z)=0$. However, in the case of a QW heterostructure, at the in-plane $\Gamma$ point of the in-plane Brillouin zone $\vec{k}=(k_y,k_z)=0$ but the momentum along the growth direction is not defined ($\hat{k}_x \ne k_x$) and in particular can never be exactly zero for any bound state. Additionally, the larger $|k_x|$ components will be increasingly present in bound states as the width of the QW reduces. Thus the inter-band conduction–valence mixing is always present in such system due to quantum size effect.

Again, the material parameters for InAs and GaSb are taken from Ref.~\cite{vurgaftman2001band} and given in Appendix~\ref{sec:material_parameters}. For the $8$~nm QW, $10$ electron wavefunctions and $8$ hole ones are used in the computation. The output energies $E_{n,k}^d$ and wavefunction $\psi_{n,k}^d$ were calculated on $k$-space mesh with $51$ mesh points along each of the directions ${d \in \lbrace [10], [11] \rbrace}$ and mesh spacing $\Delta_k = 0.3$~nm$^{-1}$.

One of the potential problems of the $8\times8~\vec{k}\cdot\vec{p}$ numerical algorithm implementation in the case of heterostructures is the existence of so called spurious solutions, see the extended discussion in Refs.~\cite{andlauer2004discretization,birner2006modeling,ehrhardt2014multiband}. In order to avoid this problem, the $S=1$ renormalization is used, which is the default implementation in \textit{nextnano++}. This is equivalent to attributing the whole modification of the conduction band curvature at the $\Gamma$ point to the conduction-valence coupling, as opposed to contribution from the far bands, or equivalently, putting the most right hand side term in equation Eq.~(\ref{eq:S_renormalization}) to zero.

\begin{figure*}[hbt!]
    \centering
    \includegraphics[width=0.33\linewidth]{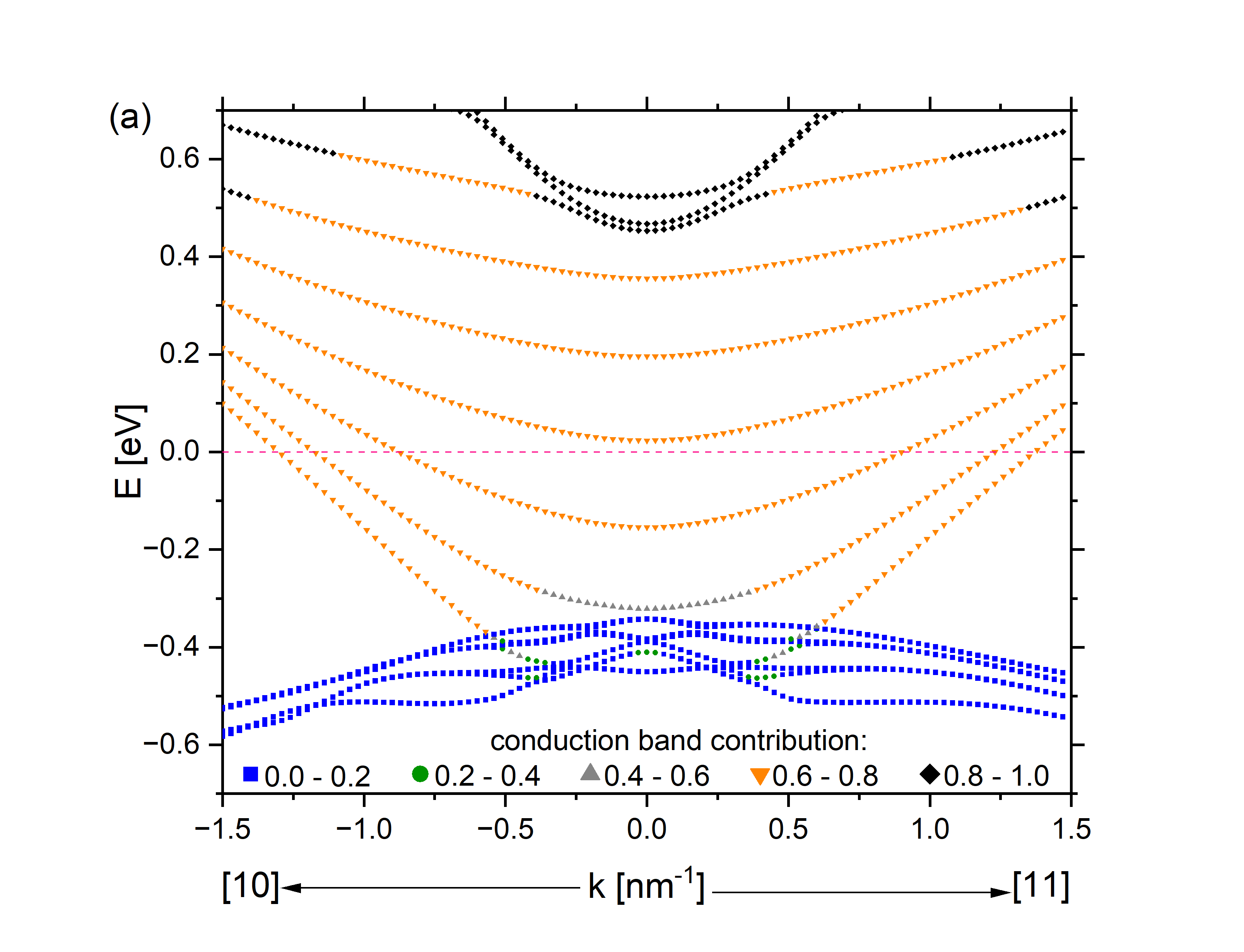}
    \includegraphics[width=0.33\linewidth]{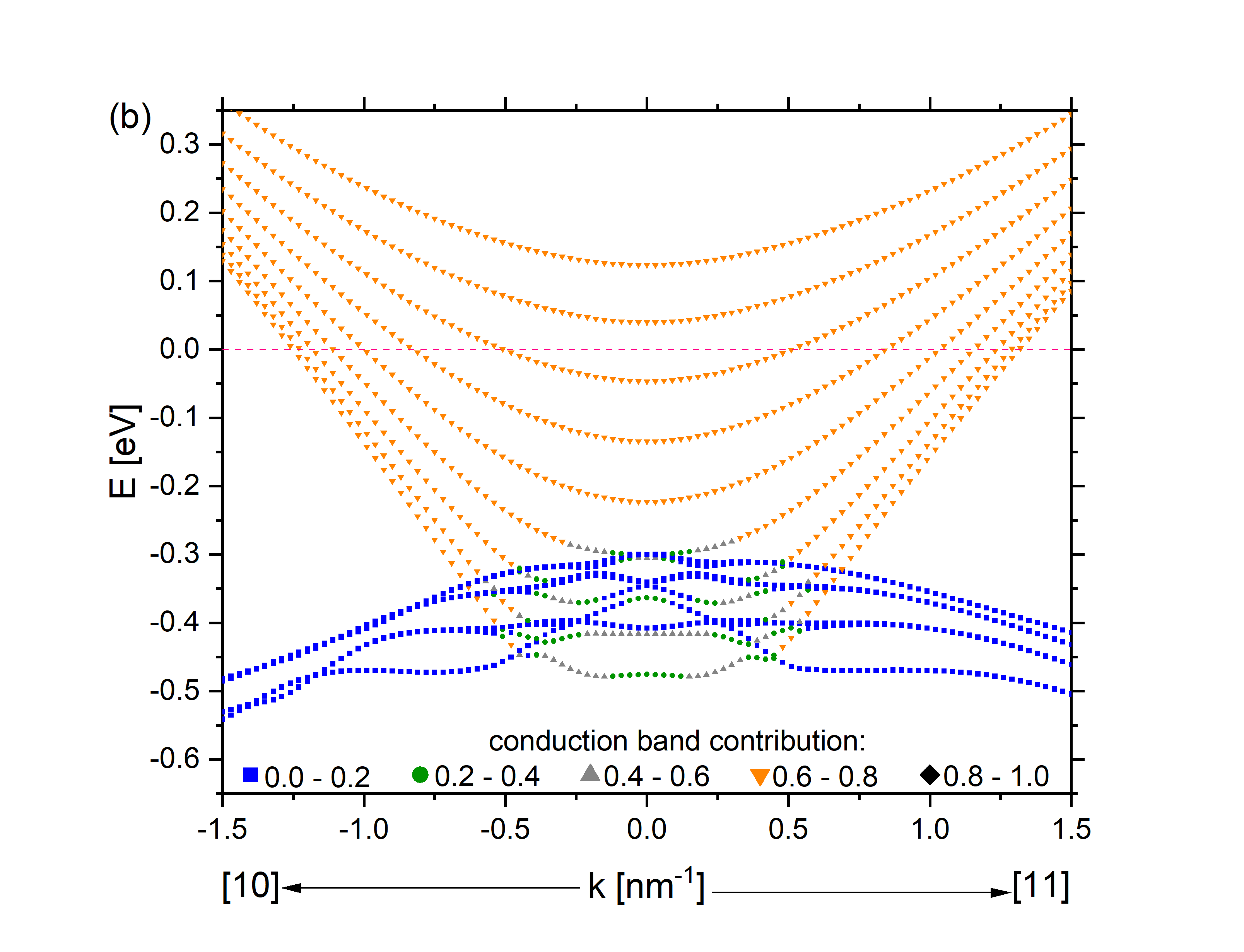}
    \caption{Color mapped dispersions showing the conduction Bloch bands contribution of the \ekdp eigenstates (a)~for the $8$~nm QW system, (b)~for the $18$~nm QW system. The dashed horizontal line at zero energy corresponds to the Fermi level.}
    \label{fig:dispersion_composition_color_maps}
\end{figure*}

Figures \ref{fig:8nm_dispersion_comparison}(b) and \ref{fig:18nm_dispersion_comparison}(b) show the subband dispersion of $8$~nm QW and $18$~nm QW systems, respectively, obtained from the \ekdp, with the Fermi level taken as the energy reference. It should be noted that the energy separations predicted by the \ekdp are visibly smaller than the corresponding ones obtained from the single-band Hamiltonian, Figs.~\ref{fig:8nm_dispersion_comparison}(a) and \ref{fig:18nm_dispersion_comparison}(a). The separations can be directly compared between the figures, as the energy scale is intentionally kept the same. This is a crucial difference between the models, which qualitatively affects the predictions of the plasmon models, as explained in Sec.~\ref{sec:semiclassical_plasmon_models}. This difference comes from the different types of the single-particle Hamiltonians of both models. In the single-band Hamiltonian, the effective mass parameter of the conduction band is obtained directly from the second order perturbation theory. Conversely, in the \ekdp there are two groups: coupling between the $8$ Bloch bulk bands close to Fermi level is taken explicitly and only remote bands are accounted for perturbatively. It is well known that the simple perturbation theory fails for near-degenerate levels and cannot properly describe the formation of avoided crossings which can be only obtained via the direct diagonalization of the Hamiltonian in the basis of relevant levels.

To illustrate the qualitative impact of the mixing, the color maps in Figs.~\ref{fig:dispersion_composition_color_maps}(a) and (b) show the contribution of the conduction bulk Bloch bands to the $8$-band eigenvector, with the total contribution of the valence bands equal to one minus the former quantity. At low energy, anti-crossing points of electron and hole energy levels are formed in the energy spectrum due to mixing between valence and conduction quantum states (with the gray dots corresponding to the states of approximately equal conduction and valence contributions) and the corresponding strong hybridization of the orbitals takes place. The Fermi level is energetically separated by at least $280$~meV from the mentioned maximum mixing range for the both systems. However, even the states in the Fermi level vicinity have between $20\%$ and $40\%$ of the total bulk valence band contributions, see the orange dot color in Figs.~\ref{fig:dispersion_composition_color_maps}(a) and (b). Since the states lying directly below the Fermi level are the highest-energy occupied states and the ones directly below the Fermi level are the lowest-lying empty states, these will play the most significant role in the formation of the plasmon modes. The large contribution of the bulk valence bands to these QW subband states validates the assumption that the inter-band mixing dynamic may turn out to be qualitatively meaningful in the further analysis, as will be explicitly demonstrated in the sections below.

Note that the individual dots in Figs.~\ref{fig:dispersion_composition_color_maps}(a) and (b) correspond to the mesh of $\vec{k}$ points used in the plasmon \ekdp discussed in Sec.~\ref{sec:ekdp_model}, as shown in Fig.~\ref{fig:k_mesh_scheme}.

The eigenfunctions of the $8$-band Hamiltonian are doubly Kramer degenerate: each pair of levels corresponds to the same orbital but with opposite spin chiralities. Due to this degeneracy, any linear combination of the states involved can be numerically obtained as a solution to the Schrödinger-Poisson equation. The procedure to de-hybridize the degenerate eigenfunctions and classify them into two groups of two opposite spin chiralities is discussed in Appendix~\ref{sec:De-hybridization_of_the_spin_chiralities}.

\vspace{12pt}\section{Semiclassical plasmon models}\label{sec:semiclassical_plasmon_models}

\subsection{Parabolic model}\label{sec:parabolic_model}

\begin{figure*}[hbt!]
    \centering
    \includegraphics[width=0.33\linewidth]{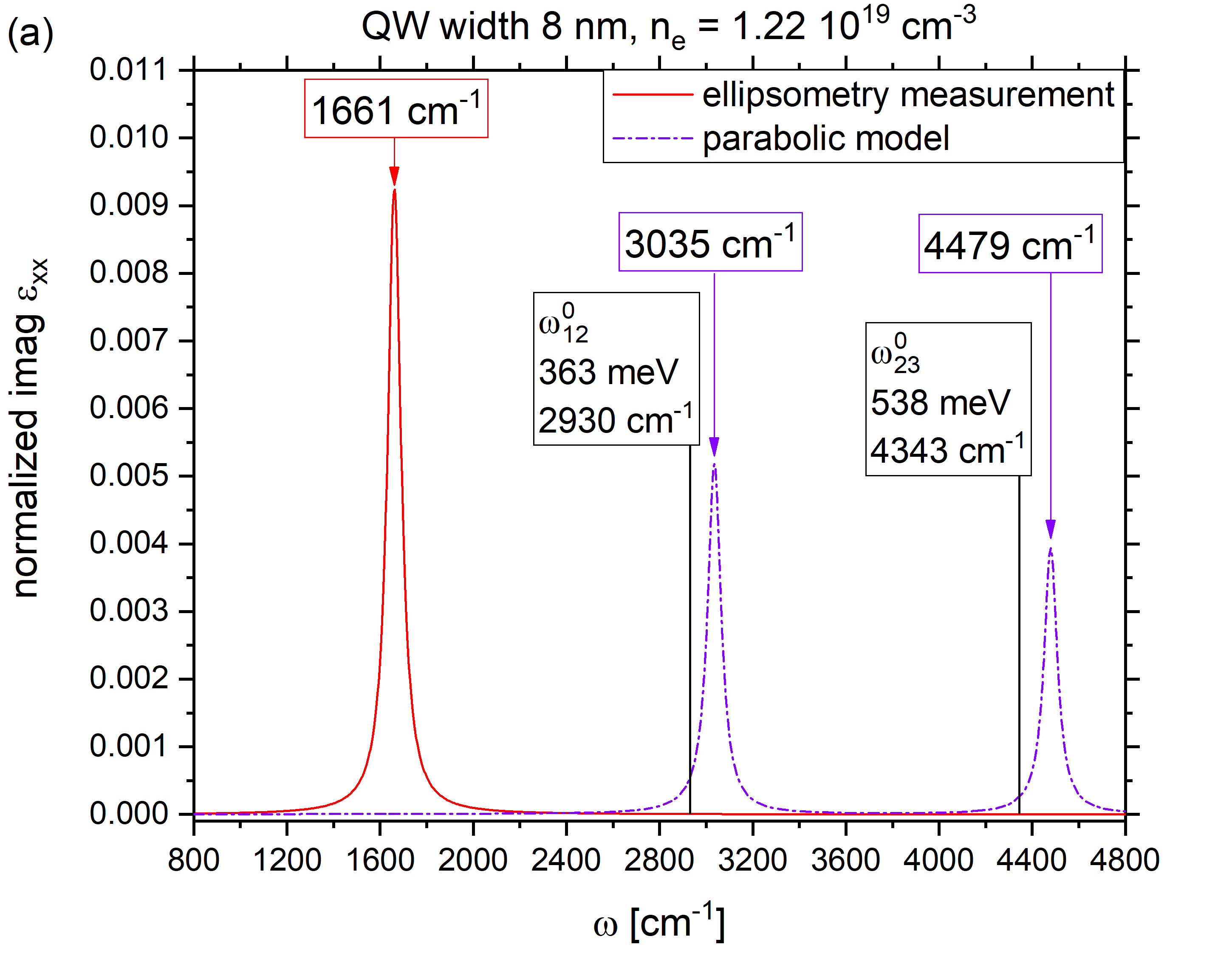}
    \includegraphics[width=0.33\linewidth]{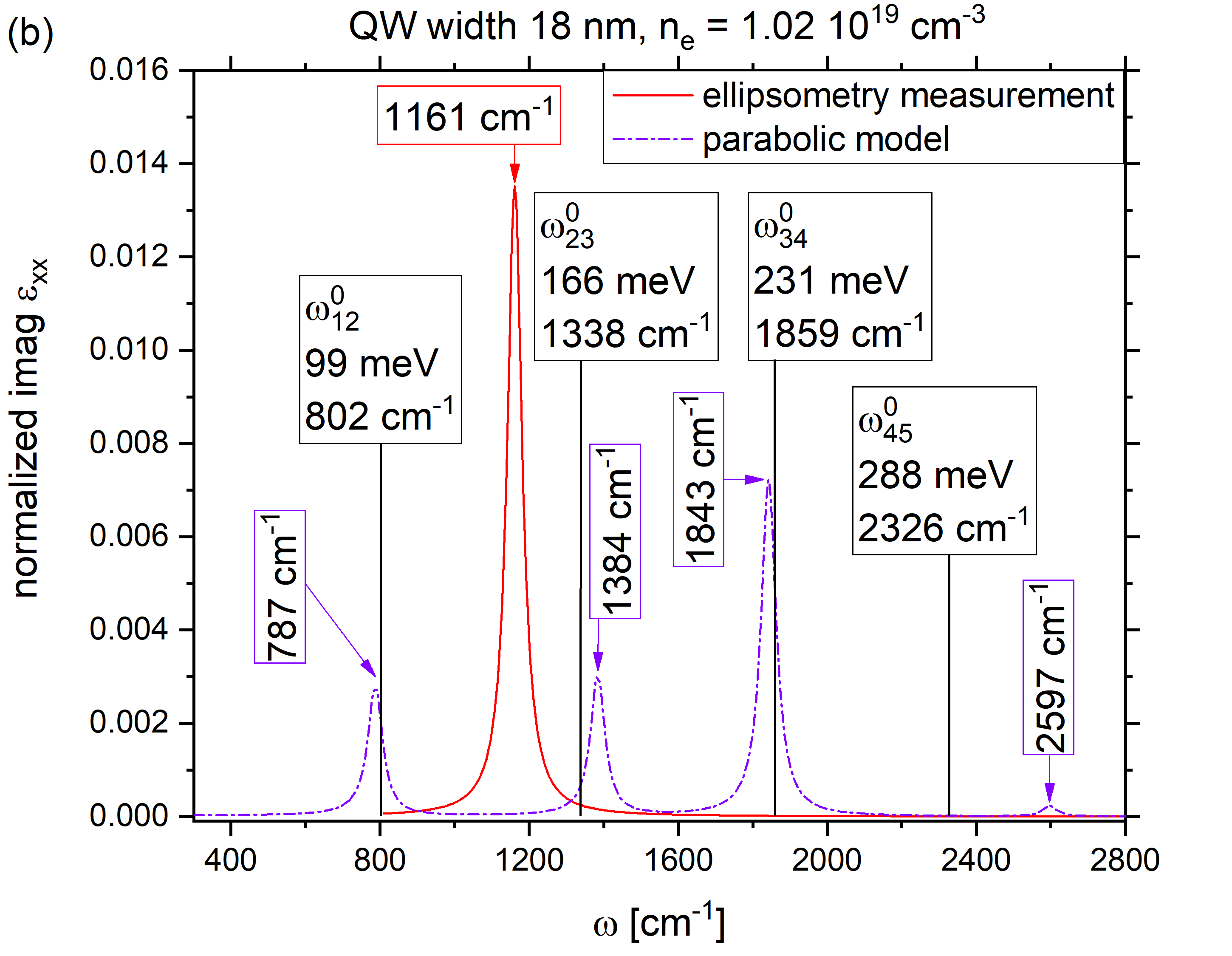}
    \caption{Detailed analysis of the imaginary part of the effective dielectric function $\epsilon_{xx}$ for the parabolic model (violet dash-dotted line). Results shown in (a) are for the $8$~nm QW system, the ones shown in (b) correspond to the $18$~nm QW one. The red solid line shows the ellipsometry measurement for comparison. Vertical black lines show the single-electron transition energies at the $\Gamma$ point $\omega_\alpha^0$. Spectra are normalized.}
    \label{fig:parabolic_analysis}
\end{figure*}

\subsubsection{General description of the parabolic model}

The simplest semi-classical approach to describe the formation of MSP modes in QWs is the parabolic model. This model was used in Ref.~\cite{alpeggiani2014semiclassical}, but will be summarized below for ease of reference.

The main idea of this approach to describe the response to an electrodynamic stimulus is to employ the non-local electric susceptibility $\chi(\omega,\vec{r},\vec{r'})$, defined by the equation
\begin{equation}
    \vec{P}(\omega,\vec{r}) = \int \chi(\omega,\vec{r},\vec{r'}) \vec{E}(\omega,\vec{r'}) d^3\vec{r'}
\end{equation}
where $\vec{P}(\omega,\vec{r})$ is the polarization response at point $\vec{r}$ to the electric field stimulus $\vec{E}(\omega,\vec{r'})$ of frequency $\omega$ at point $\vec{r'}$ and to characterize the plasmon modes of the system with this quantity. In the scope of the effective mass approximation, only the component in the growth direction $x$ is relevant. For direct transitions, it can be expressed (see Ref.~\cite{eguiluz1978electromagnetic}) as
\begin{equation}
    \chi_{xx}(\omega,x,x') = \sum_\alpha \chi_\alpha(\omega) \xi_\alpha(x) \xi_\alpha(x'),
\end{equation}
Here, $\alpha = (n_{I},n_{F})$ is a quantum number referring to the transition between the initial $n_{I}$ and final $n_{F}$ states. The $\chi_\alpha(\omega)$ part is the susceptibility for a given transition $\alpha$, which characterizes the impact of the dispersion relation (via difference in energy $\omega_\alpha(\vec{k})$ and in occupation ${\Delta}f_{\alpha,\vec{k}}$ between the levels involved) on the strength of the transition, while the inter-subband current element $\xi_\alpha(x)$ characterizes the impact of the symmetries of the initial and final wavefunctions. The former has the form
\small\begin{equation}
    \chi_\alpha(\omega) = \frac{-\hbar e^2}{\omega^2 \epsilon_0 \omega_\alpha^0 S {m^{*}}^2} \sum_{\vec{k}} {\Delta}f_{\alpha,\vec{k}} \left( 1 + \frac{\omega_\alpha(\vec{k}) \omega_\alpha^0}{(\omega+i\eta)^2-\omega_\alpha^2(\vec{k})} \right),\label{eq:ISP_susceptibility}
\end{equation}\normalsize
where $\omega_\alpha^0 = \omega_\alpha(\vec{k}=0)$, $S$ is the surface and $\eta$ is the homogeneous broadening parameter. On the other hand, the inter-subband current matrix element is given by
\begin{equation}
    \xi_\alpha(x) = \psi_{n_{I}}(x) \frac{\partial \psi_{n_{F}}(x)}{\partial x} - \frac{\partial \psi_{n_{I}}(x)}{\partial x} \psi_{n_{F}}(x),
\end{equation}
with the $\psi_{n}(x)$ representing the single-particle solutions of the Schrödinger-Poisson equation.
Note that, already at this stage, it was assumed in the model that the wavefunctions $\psi_{n}(x)$ do not depend not-trivially on the wavevector $\vec{k}$. Moreover, the dependence of the effective mass $m^{*}$ on the type of the material (well vs barrier) and thus on the position $x$ was not taken into account. Since most of the electron density, in the case of the conduction band, is localized inside the well, the ${m_{\text{InAs}}^{*}} = 0.026.m_{0}$ value is used in the parabolic model. Furthermore, if also the energy separation is constant in the whole Brillouin zone, as in the case of a simple isotropic and parabolic in-plane dispersion relation, then $\omega_\alpha^0 \equiv\omega_\alpha (\vec{k})$
and
\begin{equation}
    \chi_\alpha(\omega) = -\frac{\hbar e^2 {\Delta}n_{\alpha}}{2 \epsilon_0 \omega_\alpha {m^{*}}^2} \frac{1}{(\omega+i\eta)^2-\omega_\alpha^2}.
\end{equation}
where ${\Delta}n_\alpha$ is the 2D population density difference between subbands $n_I$ and $n_F$. In this case, the squared resonant frequencies of the MSP mode $\Omega^2$ will be the eigenvalues of the $M$ matrix representing the coupling of the individual ISP modes, as follows
\begin{equation}
    M_{\alpha,\alpha'} = \omega_\alpha^2 \delta_{\alpha,\alpha'} + {\omega_P}_\alpha {\omega_P}_{\alpha'} \frac{I_{\alpha,\alpha'}}{\sqrt{I_{\alpha,\alpha}I_{\alpha',\alpha'}}},
\label{eq:M_matrix}\end{equation}
where the plasma frequencies of the individual ISP modes are
\begin{equation}
 {\omega_P}^2_\alpha = \frac{\hbar e^2}{2 \epsilon_0 \epsilon_S {m^*}^2} \frac{ {\Delta}n_\alpha I_{\alpha,\alpha}}{\omega_\alpha}   
\label{eq:omega_P_2_def}\end{equation}
and
\begin{equation}
    I_{\alpha,\alpha'} = \int \xi_\alpha(x) \xi_{\alpha'}(x)dx.
\label{eq:I_aa'}\end{equation}

The bulk permittivity $\epsilon_S$ can be identified with the dielectric constant of the QW material (InAs) since most of the density of the relevant eigenstates is localized inside the well. Two simplifying assumptions are made here. Firstly, the parabolic model treats $\epsilon_S$ as uniform, neglecting the (relatively small) difference between the dielectric properties of InAs and GaSb. Secondly, the $\epsilon_S$ is assumed to be independent of the frequency $\omega$, however it is known that the electrostatic dielectric constant is somewhat different from the one corresponding to the high frequencies (about $21\%$ in the case of InAs). The corresponding values for the materials in question are included in Table~\ref{tab:material_parameters} in Appendix~\ref{sec:material_parameters}.

Finally, the non-local susceptibility for the whole MSP takes the form
\begin{equation}
    \chi_{xx} = -\sum_j \frac{\beta_j^2}{\omega^2 - (\Omega_j-\frac{i}{2} \gamma_j)^2}, 
\end{equation}
where $j$ numerates the MSP modes, the effective peak broadening parameter is $\gamma_j$ (in this work always fitted to the experimental data) and the amplitudes are given by
\begin{equation}
    \beta_j^2 = \left( \sum_\alpha {\omega_P}_\alpha \varepsilon_{\alpha,j} \frac{\zeta_\alpha}{\sqrt{I_{\alpha,\alpha}}} \right)^2,
\label{eq:beta_j_def}\end{equation}
where $\varepsilon_{\alpha,j}$ is the $\alpha$-th ISP component of the $j$-th MSP mode eigenvector of the $M$ matrix and
\begin{equation}
    \zeta_\alpha = \int \xi_\alpha(x) dx
\label{eq:inter_subband_dipole_moment}\end{equation}
stands for the inter-subband transition moment.
For the convenience of the future modification of the model, the quantities will be expressed with the ${\omega_P}_\alpha$ expanded:
\begin{align}
    M_{\alpha,\alpha'} &= \omega_\alpha^2 \delta_{\alpha,\alpha'} + \frac{\hbar e^2}{2 \epsilon_0 \epsilon_S} \sqrt{\frac{{\Delta}n_\alpha}{\omega_\alpha}} \sqrt{\frac{{\Delta}n_\alpha'}{\omega_\alpha'}} \frac{I_{\alpha,\alpha'}}{{m^*}^2},\\
    \beta_j^2 &= \frac{\hbar e^2}{2 \epsilon_0 \epsilon_S} \left( \sum_\alpha \varepsilon_{\alpha,j} \sqrt{\frac{{\Delta}n_\alpha}{\omega_\alpha}} \frac{\zeta_\alpha}{m^*} \right)^2.
\end{align}

Note that in the single-band Hamiltonian of Sec.~\ref{sec:single-band_single_particle}, the subbands are doubly degenerate due to spin. Since the spin conserving transitions are allowed but the ones between the opposite spins are forbidden, the spin can be disregarded in the calculation, with the simultaneous doubling of the 2D population density difference between subbands ${\Delta}n_\alpha$.

\subsubsection{Effective local permittivity}

In the scope of the effective medium approach, the effective electric permittivity (along the growth direction~$x$) can be defined as
\begin{equation}
    \tilde{\epsilon}_{xx} = \frac{\left\langle D_x \right\rangle}{\epsilon_0 \left\langle E_x \right\rangle},
\end{equation}
where the $D_x$, $E_x$ are the corresponding components of the displacement and the electric fields, respectively. Under the additional assumptions that the bulk permittivity $\epsilon_S$ is the same as the in-plane permittivity of the heterostructure $\epsilon_{||}$, and that the averaging $\left\langle \ldots \right\rangle$ is done exactly over the width of the nanostructure $W_{\text{eff}} = 2 W_{\text{GaSb}} + W_{\text{InAs}}$, the latter expression simplifies to
\begin{equation}
    \tilde{\epsilon}_{xx} = \frac{\left\langle D_x \right\rangle}{\epsilon_0 \left\langle E_x \right\rangle} \approx \epsilon_S \left[1 - \frac{\chi_{xx}}{W_{\text{eff}}} \right]^{-1},
\label{eq:chi_to_epsi}\end{equation}
thus the effective dielectric function can be directly obtained from the susceptibility $\chi_{xx}$.
The full derivation can be found in Ref.~\cite{pasek2022multisubband}.

\subsubsection{Simulation results of the parabolic model}

Fig.~\ref{fig:parabolic_analysis} illustrates the $\Im(\epsilon_{xx})$ spectrum obtained with the parabolic model for $8$~nm QW in (a) and for $18$~nm QW in (b). In the former case, the parabolic model (violet dash-dotted line) predicts two peaks: one at $3035$\rcm and the other at $4479$\rcm. These predictions differ qualitatively from the experimental result of a single MSP peak at $1661$\rcm, as shown by the red solid line.
In fact the two peaks of the parabolic model are not the MSP peaks, but individual uncoupled ISP. As an illustration, the two black vertical lines in Fig.~\ref{fig:parabolic_analysis}(a) show the IST transition frequencies $\omega_\alpha^0$ for transitions $1\rightarrow2$ at $2930$\rcm and $2\rightarrow3$ at $4343$\rcm, which correspond to the energy separation of the corresponding subbands at the $\Gamma$ points.
It is visible that each of the plasmon peaks is just a slightly shifted IST. These IST$\rightarrow$ISP plasmon shifts ($105$\rcm and $136$\rcm, respectively) are small in comparison to their separation of $\omega_{23}^0 - \omega_{12}^0 = 1413$\rcm, the latter one being a direct consequence of the single-particle spectrum shown in Fig.~\ref{fig:8nm_dispersion_comparison}(a).
In the case of the $18$~nm system, shown in Fig.~\ref{fig:parabolic_analysis}(b), a very similar situation happens for the transitions $1\rightarrow2$, $2\rightarrow3$, and $3\rightarrow4$: the plasmon peaks almost coincide with the IST energies.
The plasmon shift is bigger in the case of $\omega_{45}$, but this is a very small peak with negligible impact on the character of the whole spectrum.
In the ellipsometry spectroscopy, a single MSP peak at $1161$\rcm was found instead (red solid line) in the case of this system.
To sum up, the parabolic model does not predict the formation of main MSP peak in the spectrum, but almost uncoupled ISP peaks, in contrast to what was observed in the ellipsometry experiment. As was explained before, this difference arises from the limited quantum description of the QW, because the single-band Schrödinger-Poisson equation does not account for non-parabolicity and mixing of the conduction and valence quantum states due to quantum size effect.

\subsection{Hybrid model}\label{sec:hybrid_model}
\begin{figure*}[hbt!]
    \centering
    \includegraphics[width=0.33\linewidth]{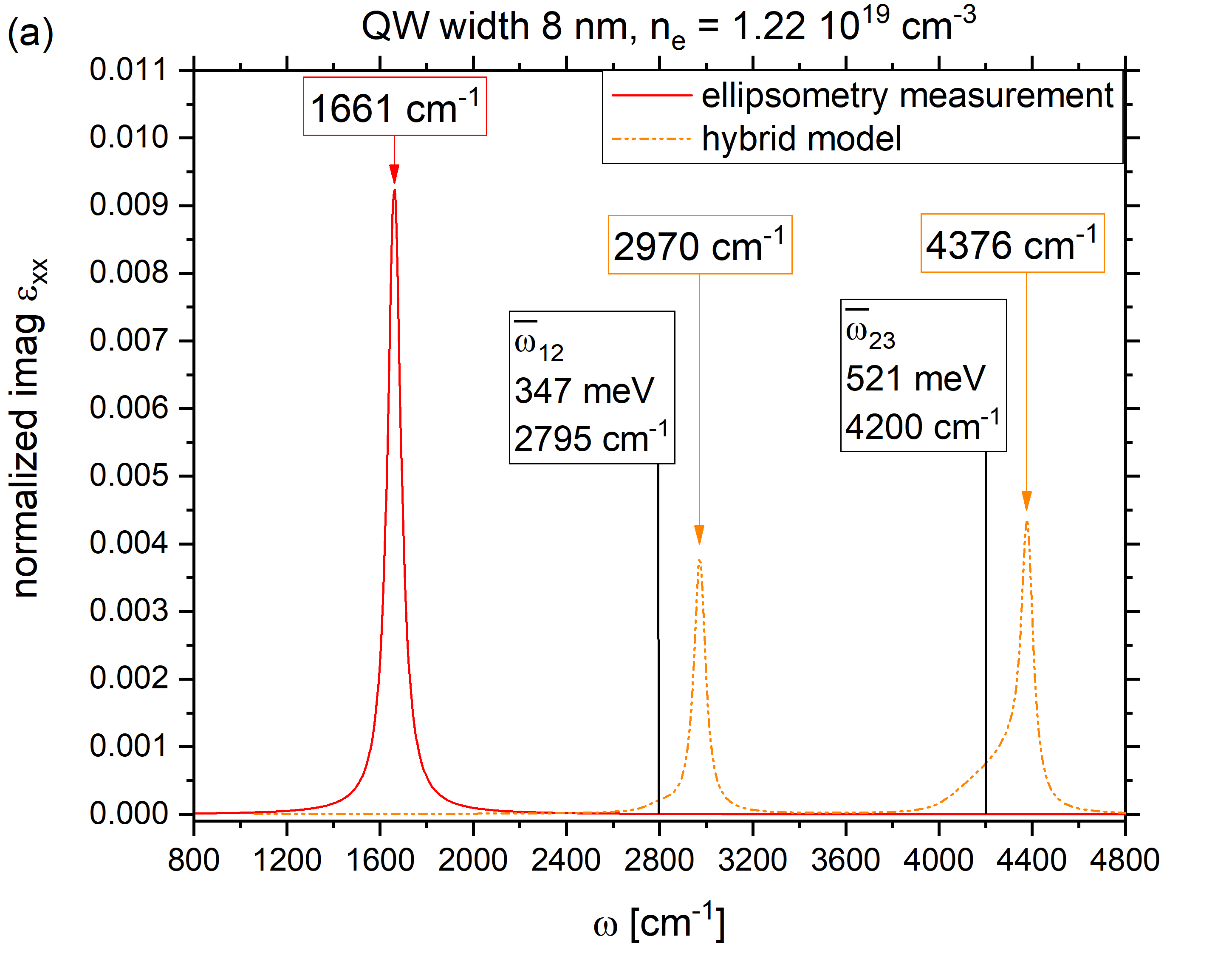}
    \includegraphics[width=0.33\linewidth]{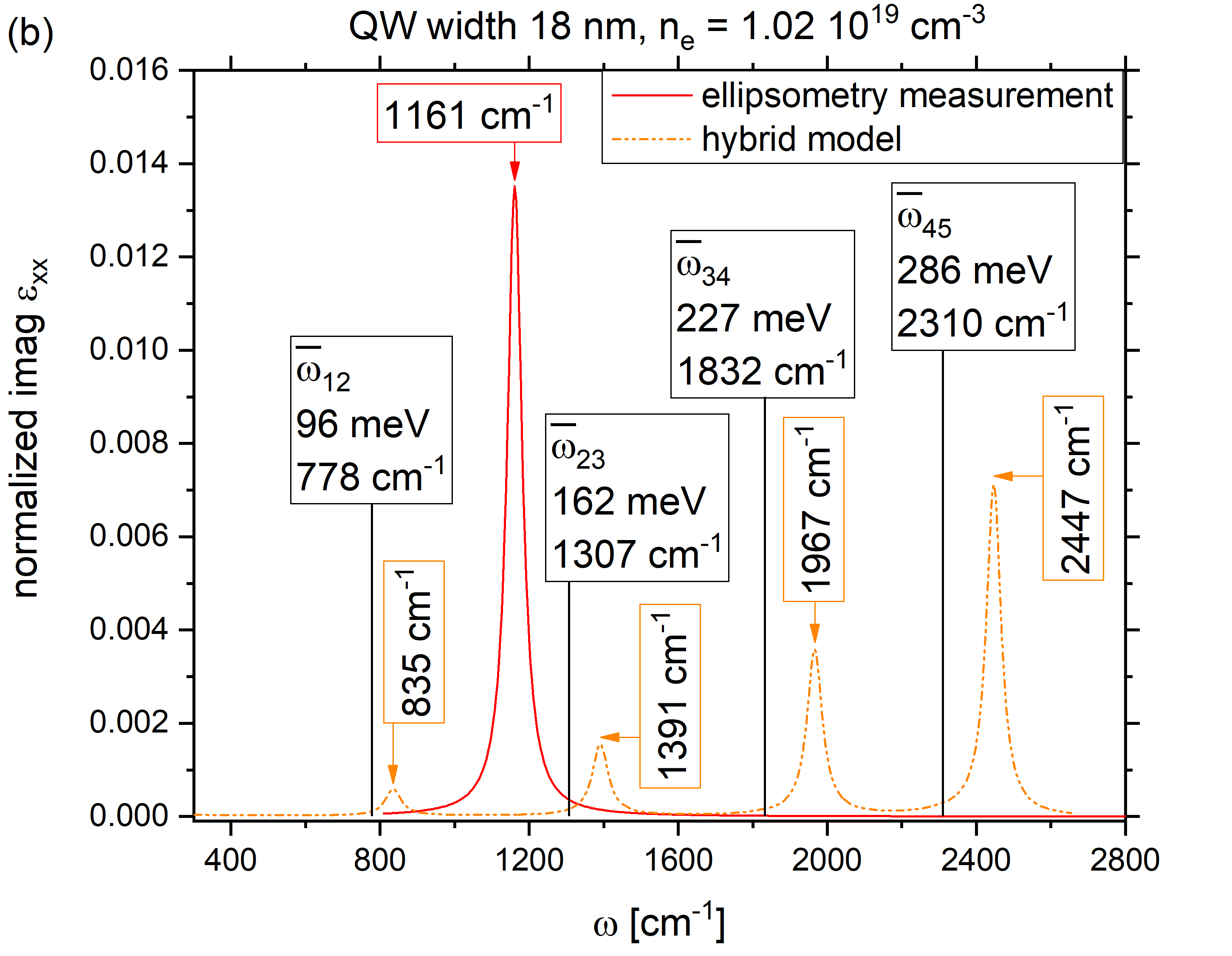}
    \caption{Detailed analysis of the imaginary part of the effective dielectric function $\epsilon_{xx}$ for the hybrid model (orange dash-dot-dotted line). Results shown in (a) are for the $8$~nm QW system, the ones shown in (b) correspond to the $18$~nm QW one. The red solid line shows the ellipsometry measurement for comparison. Vertical black lines show $\overline{\omega_{\alpha}} = \sqrt{\overline{\omega_{\alpha}^2}}$, the single-electron mean transition energies averaged over $\vec{k}$, compare with Eq.~(\ref{eq:mean_omega_alpha}). Spectra are normalized.}
    \label{fig:hybrid_analysis}
\end{figure*}

\subsubsection{General description of the hybrid model}

The hybrid model takes into account the $\vec{k}$-dependence of the $\omega_\alpha(\vec{k})$, resulting from non-parabolicity of the in-plane subband dispersion relation. It is worth mentioning that this non-parabolicity is not due to bulk dispersion of the conduction bands, which are isotropic and parabolic in the single-band model, but due to non-homogeneity of the heterostructure. The latter is composed of two different materials in different $x$ regions, which have two different dispersion relations, imposing the quantum size effect on the whole nanosystem. In effect, the dispersion relations are isotropic but depend on the magnitude of $\vec{k}$. Thus, one can proceed directly from the Eq.~(\ref{eq:ISP_susceptibility}), changing sum over states to an integral over the relevant part of Brillouin zone, integrating over angle but keeping the $|\vec{k}|$-dependence and then finally discretizing the momentum mesh ($k \rightarrow k_i$, ${\Delta}_k = k_{i+1} - k_i$) to obtain the effective sum:

\footnotesize\begin{equation}
    \chi_\alpha(\omega) = \frac{-\hbar e^2 {\Delta}_k}{2\pi \epsilon_0 {m^*}^2 \omega_\alpha^0 \omega^2} \sum_i k_i {\Delta}f_{\alpha,k_i} \left(1 +\frac{ \omega_{\alpha}(k_i) \omega_\alpha^0}{(\omega + \frac{i\gamma}{2})^2 - \omega_{\alpha}(k_i)^2}\right)
\end{equation}\normalsize

In this case, the constituting coefficients of the susceptibility are given by the vector $F$, which can be obtained by solving the set of linear equations $L{\cdot}F=R$, where the left hand side matrix and the right hand side vector are given, respectively, by
\begin{equation}
 L_{\alpha,\alpha'} = \delta_{\alpha,\alpha'} + \frac{\chi_{\alpha'}(\omega)}{\epsilon_S} I_{\alpha,\alpha'},~R_\alpha = \zeta_\alpha.  
\end{equation}
Having obtained the $F_\alpha$ coefficients, one can calculate the susceptibility as
\begin{equation}
    \chi_{xx} = \frac{1}{\epsilon_S} \sum \chi_\alpha(\omega) F_\alpha \zeta_\alpha.
\end{equation}

As was explained before in the case of the parabolic mode, spin can be disregarded in the hybrid model if the 2D population density difference between subbands ${\Delta}n_\alpha$ is multiplied by two.

\subsubsection{Simulation results of the hybrid model}

Figure~\ref{fig:hybrid_analysis}(a) illustrates the $\Im(\epsilon_{xx})$ spectrum of the $8$~nm system. The hybrid model predicts two peaks: one located at $2970$\rcm and the other at $4376$\rcm, as shown by the orange dash-dot-dotted line. The solid red line corresponds to the experimental observations exhibiting a single MSP peak at $1661$\rcm. 
In order to compare the plasmon peaks predicted by the hybrid model to the ISTs, a new parameter characterizing the latter needs to be used, as the energy separations $\hbar \omega_\alpha(k)$ depend now on $k = |\vec{k}|$. It is reasonable to use an average of the $\omega_\alpha(\vec{k})$ weighted by the occupation difference of the initial and final states involved $\overline{\omega_\alpha} = \sqrt{\overline{\omega_\alpha^2}}$, see Eq.~(\ref{eq:mean_omega_alpha}). As was the case in the parabolic model, the two peaks in the hybrid model correspond to slightly shifted IST frequencies $\overline{\omega_{12}}$ and $\overline{\omega_{23}}$, as shown by the black vertical lines in the figure. Also in this case, the plasmon energy shifts are small in comparison to the energy separation of the peaks, thus pointing to the single-particle model as the cause of this discrepancy.
The results in Fig.~\ref{fig:hybrid_analysis}(b) show the $\Im(\epsilon_{xx})$ spectrum of the $18$~nm system. Again, the situation for the hybrid model is qualitatively similar to the parabolic model. That is to say, there is no single MSP peak predicted by the model (in contrast to the experimental one at $1161$\rcm) but instead the simulated spectrum show series of ISP peaks each approximately corresponding to an individual IST frequency: $\bar{\omega}_{12}$, $\bar{\omega}_{23}$, $\bar{\omega}_{34}$ and $\bar{\omega}_{45}$, respectively.

\subsection{Generalization of the semiclassical model to the \ekdp case}\label{sec:ekdp_model}

\subsubsection{General description of the \ekdp}

In order to generalize the Alpeggiani-Andreani model of Ref.~~\cite{alpeggiani2014semiclassical} to the non-parabolic scenario, the following four points should be taken into account.

Firstly, the eigenfunction $\Psi_{n,\vec{k}}^\mathcal{B}(x)$ of the Hamiltonian will now be an $8$-element vector written in the $\ket{J,J_z}$ Bloch basis of $s$-type conduction band $\ket{S\frac{1}{2},\pm\frac{1}{2}}$, and three $p$-type valence bands: heavy hole $\ket{\frac{3}{2},\pm\frac{3}{2}}$, light hole $\ket{\frac{3}{2},\pm\frac{1}{2}}$ and split-off bands $\ket{\frac{1}{2},\pm\frac{1}{2}}$. This basis is a natural choice, as the bulk (no strain or quantum size effect) Hamiltonian is diagonal at the $\Gamma$ point in this basis. Note that the effective mass $m^\mathcal{B}(x)$ is different for each Bloch band $\mathcal{B}$.

Secondly, each of the Bloch band's effective masses $m^\mathcal{B}(x)$ as well as the electrical permittivity $\epsilon_S(x)$ are different in the well material InAs from the corresponding ones of the barrier material GaSb. In the model of Ref.~\cite{alpeggiani2014semiclassical} these are considered homogeneous, but here they are functions of the position along the out-of-plane axis $x$, which means they should be absorbed inside the integration over real space in Eqs.~(\ref{eq:I_aa'}) and (\ref{eq:inter_subband_dipole_moment}). Note that this eliminates one of the two simplifications about $\epsilon_S$ done previously, but the dependence on $\omega$ is still not considered. In the \ekdp, the values of $\epsilon_S = 12.30~\epsilon_0$ and $\epsilon_S = 14.40~\epsilon_0$ for InAs and GaSb are adopted (see Table~\ref{tab:material_parameters} in Appendix~\ref{sec:material_parameters}), respectively, corresponding to the high frequencies, as the characteristic frequencies of the ISP and MSP lie in this frequency range.

Thirdly, some other quantities considered in Ref.~\cite{alpeggiani2014semiclassical} as constant over the whole Brillouin zone, now depend on both the magnitude and the orientation of the wavevector $\vec{k}$. Explicitly, they are: $\xi_\alpha(x) \rightarrow \xi_{\alpha,\vec{k}}(x)$, $I_{\alpha,\alpha'} \rightarrow I_{\alpha,\vec{k},\alpha',\vec{k}'}$ and $\omega_{\alpha} \rightarrow \omega_{\alpha}(\vec{k})$. Consequently, they should be included inside the integration over $\vec{k}$ in Eq.~(\ref{eq:ISP_susceptibility}).
Because the eigenfunctions $\Psi_{n,\vec{k}}^\mathcal{B}(x)$ are non-trivially complex, some of other quantities are such as well. Coherent conjugation rules should be used, as  $\xi_{\alpha,\vec{k}}(x) \ne {\xi_{\alpha,\vec{k}}(x)}^{*}$ and $I_{\alpha,\vec{k},\alpha',\vec{k}'} \ne {I_{\alpha,\vec{k},\alpha',\vec{k}'}}^{*}$.

Regarding the transition $\alpha: n_{I} \rightarrow n_{F}$, one should start with defining:
\small\begin{multline}
    \tilde{\xi}_{\alpha,\vec{k}}(x) = \frac{1}{\sqrt{\epsilon_S(x)}} \sum_{\mathcal{B}} \left( \Psi_{n_{I},\vec{k}}^\mathcal{B}(x) \frac{1}{m^\mathcal{B}(x)} \frac{\partial}{\partial x} {\Psi_{n_{F},\vec{k}}^\mathcal{B}(x)}^{*} \right.\\
    \left. - {\Psi_{n_{F},\vec{k}}^\mathcal{B}(x)}^{*} \frac{1}{m^\mathcal{B}(x)} \frac{\partial}{\partial x} \Psi_{n_{I},\vec{k}}^\mathcal{B}(x) \right).
\label{eq:def_tilde_xi_alpha_k}\end{multline}\normalsize
The values of the effective masses correspond to the ones of uncoupled bands in the direction of the spin projection, that is
\begin{align}
\frac{1}{m^{\left\lvert S,\pm 1/2 \right\rangle}(x)} &= 1/m_e^{*} \nonumber\\
\frac{1}{m^{\left\lvert \frac{3}{2},\pm \frac{3}{2} \right\rangle}(x)} &= - \left(\gamma^{6 \times 6}_1 - 2 \gamma^{6 \times 6}_2\right) \nonumber\\
\frac{1}{m^{\left\lvert \frac{3}{2},\pm \frac{1}{2} \right\rangle}(x)} &= - \left(\gamma^{6 \times 6}_1 + 2 \gamma^{6 \times 6}_2\right) \nonumber\\
\frac{1}{m^{\left\lvert \frac{1}{2},+\frac{1}{2} \right\rangle}(x)} &= -\gamma^{6 \times 6}_1,
\label{eq:def_Bloch_band_masses}\end{align}
where $\gamma^{6 \times 6}$ are the \textit{unmodified} Luttinger parameters and the $x$ dependence exists in the form of corresponding to either InAs or GaSb. This way one naturally gets:
\begin{equation}
    \tilde{I}_{\alpha,\vec{k},\alpha',\vec{k}'} = \int \tilde{\xi}_{\alpha,\vec{k}}(x) {\tilde{\xi}_{\alpha',\vec{k}'}(x)}^{*} dx.
\end{equation}

\begin{figure}[hbt!]
    \centering
    \includegraphics[width=0.75\linewidth]{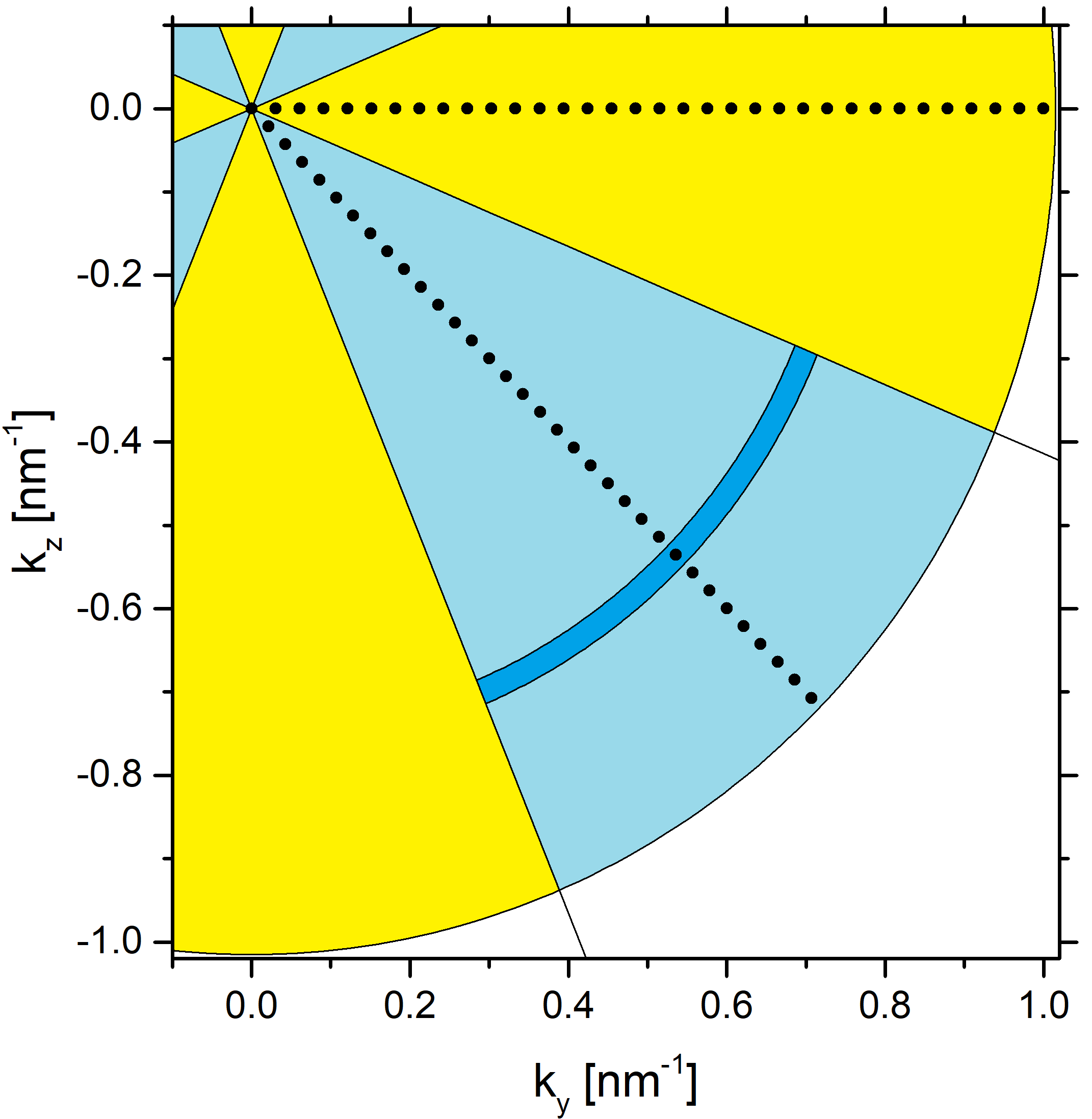}
    \caption{The scheme of the momentum space mesh. The light blue and yellow colors mark the two angular subsets. The dots represent the mesh points. The dark blue color show the area corresponding to one of the points.}
    \label{fig:k_mesh_scheme}
\end{figure}

The 2D occupation difference $\Delta n_\alpha$ of the parabolic model can be explicitly written as:
\begin{equation}
    \Delta n_\alpha = \sum_{\vec{k}} \Delta f_{\alpha,\vec{k}} = \frac{1}{4 \pi^2} \int \Delta f_{\alpha,\vec{k}} dk_z dk_y.
\label{occupation_difference}\end{equation}
In \ekdp, the relevant circle of the momentum space will be divided into two subsets, each corresponding to half of the circle, see Fig.~\ref{fig:k_mesh_scheme}. First subset (the yellow area in the figure) is represented by a $k = j_k \Delta_{k}$ mesh along one of the equivalent directions [10], [0,1], [-1,0] and [0,-1] and the other subset (the blue area in the figure) is represented by $k = j_k \Delta_{k}$ mesh running along one of the equivalent directions [11], [1,-1], [-1,1], [-1,-1]. This way, each of the $j_k$ mesh points represents an area of $\pi j_k \Delta_k$. By using a simple rectangle rule approximation for the integral in Eq.~(\ref{occupation_difference}), the explicit version for the \ekdp can be obtained as
\begin{equation}
    \Delta n_\alpha = \frac{1}{4 \pi} \sum_{j_k} j_k \Delta_k^2 \Delta f_{\alpha,\vec{k}}.
\end{equation}

In the plasmon frequency $\omega_{P,\alpha}$ as defined in Eq.~(\ref{eq:omega_P_2_def}), the sum over $j_k$ should be put explicitly in the front of quantities, which in the \ekdp will become $x$-dependent. This results in
\begin{equation}
    \omega_{P,\alpha}^2 = \sum_{j_k} \frac{1}{2} j_k \Delta_k^2 \Delta f_{\alpha,\vec{k}} \frac{1}{\omega_\alpha} \left( \frac{1}{\epsilon_S m^{*}} I_{\alpha,\alpha} \right).
\end{equation}
On the other hand, the coupling matrix of the parabolic case Eq.~(\ref{eq:M_matrix}) can be written as:
\begin{align}
M_{\alpha,\alpha'} &= \omega_{\alpha}^2 \delta_{\alpha, \alpha'} + \mathcal{D}_{\alpha,\alpha'}\\
\mathcal{D}_{\alpha,\alpha'} &=  \omega_{P,\alpha} \omega_{P,\alpha'} \frac{I_{\alpha,\alpha'}}{\sqrt{I_{\alpha,\alpha} I_{\alpha',\alpha'}}}.\label{square_quantity}
\end{align}
With this in mind, the square of the quantity of Eq.~(\ref{square_quantity}) can be written as:
\small\begin{align}
\mathcal{D}_{\alpha,\alpha'}^2 &= \omega_{P,\alpha}^2 \omega_{P,\alpha'}^2 \frac{I_{\alpha,\alpha'}^2}{I_{\alpha,\alpha} I_{\alpha',\alpha'}}\\
&= \left( \frac{\Delta_k^2}{2} \right)^2 \sum_{j_k} \sum_{j_k'} j_k j_k' \frac{\Delta f_{\alpha,\vec{k}} \Delta f_{\alpha',\vec{k}'}}{\omega_{\alpha} \omega_{\alpha'}} \left( \frac{1}{\epsilon_S {m^{*}}^2} I_{\alpha,\alpha'} \right)^2\nonumber
\end{align}\normalsize
and $\mathcal{D}_{\alpha,\alpha'}$ can be calculated as
\begin{equation}
\mathcal{D}_{\alpha,\alpha'} = \frac{\Delta_k^2}{2} \sqrt{\sum_{j_k} \sum_{j_k'} j_k j_k' \frac{\Delta f_{\alpha,\vec{k}} \Delta f_{\alpha',\vec{k}'}}{\omega_{\alpha} \omega_{\alpha'}} \left( \frac{1}{\epsilon_S {m^{*}}^2} I_{\alpha,\alpha'} \right)^2}.
\end{equation}
At this point, the substitutions $\left( \frac{1}{\epsilon_S {m^{*}}^2} I_{\alpha,\alpha'} \right)^2 \rightarrow \left\lvert \tilde{I}_{\alpha,\vec{k},\alpha',\vec{k}'} \right\rvert^2$ should be made, because $\frac{1}{\epsilon_S {m^{*}}^2}$ is already included in the term $\tilde{I}_{\alpha,\vec{k},\alpha',\vec{k}'}$, and $\omega_{\alpha} \rightarrow \omega_\alpha(\vec{k})$, $\omega_{\alpha'} \rightarrow \omega_{\alpha'}(\vec{k}')$. These substitutions represent the inclusion of the dependence of the corresponding quantities on $\vec{k}$ in the \ekdp. This leads to
\begin{equation}
\mathcal{D}_{\alpha,\alpha'} = \frac{\Delta_k^2}{2} \sqrt{\sum_{j_k} \sum_{j_k'} j_k j_k' \frac{\Delta f_{\alpha,\vec{k}}}{\omega_\alpha(\vec{k})}
\frac{\Delta f_{\alpha',\vec{k}'}}{\omega_{\alpha'}(\vec{k}')}
\left\lvert \tilde{I}_{\alpha,\vec{k},\alpha',\vec{k}'} \right\rvert^2}.
\end{equation}
It may be noticed that here the absolute value of the $\tilde{I}_{\alpha,\vec{k},\alpha',\vec{k}'}$ quantity was taken, which loses the information about its phase. This is done on purpose, as the square root is a multi-valued function in case of complex arguments, which may lead to errors. The phase information will be recovered below.

The amplitudes of the MSP modes $\beta_j$ in Eq.~(\ref{eq:beta_j_def}), in the scope of the present generalization, will take the form
\begin{align}
    \beta_j &= \sum_\alpha \mathcal{E}_{\alpha}^j \mathcal{L}_\alpha\\
    \mathcal{L}_\alpha &= \sqrt{ \omega_{P,\alpha}^2 f_\alpha L_{\text{eff},\alpha} }\nonumber\\
    &= \sqrt{\frac{\Delta_k^2}{2} \sum_{j_k} j_k \frac{\Delta f_{\alpha,\vec{k}}}{\omega_{\alpha}} \left( \int \frac{1}{\epsilon_S m^{*}} \xi_{\alpha}(x) dx \right)^2}, 
\end{align}
where advantage was taken of the fact that $\omega_{P,\alpha} > 0$. By substituting: $\left( \int \frac{1}{\epsilon_S m^{*}} \xi_{\alpha}(x) dx \right)^2 \rightarrow \left\lvert \int \tilde{\xi}_{\alpha,\vec{k}}(x) dx \right\rvert^2$ and $\omega_{\alpha} \rightarrow \omega_\alpha(\vec{k})$, one arrives at:
\begin{equation}
    \mathcal{L}_\alpha = \sqrt{\frac{\Delta_k^2}{2} \sum_{j_k} j_k \frac{\Delta f_{\alpha,\vec{k}}}{\omega_{\alpha}(\vec{k})} \left\lvert \int \tilde{\xi}_{\alpha,\vec{k}}(x) dx \right\rvert^2}.
\label{eq:L_alpha}\end{equation}
It should be noted that the information on the phase of the $\int \tilde{\xi}_{\alpha,\vec{k}}(x) dx$ has been lost, but it will be recovered later.

The generalization of all the relevant terms has been addressed at this stage, apart from the $\omega_\alpha^2 \delta_{\alpha,\alpha'}$ one. In this case, there is no natural way to include the $\vec{k}$ dependence, since there is no summation over $j_k$ to be put in front of the $\omega_{\alpha}(\vec{k})$. Instead, a form of weighted average over $\vec{k}$ will be adopted. A suggestion regarding the suitable weights can be deduced from the fact that the $\omega_\alpha^2 \delta_{\alpha,\alpha'}$ quantity is added to the $\mathcal{D}_{\alpha,\alpha'}$ one, and the latter includes the $\sqrt{\sum_{j_k} \sum_{j_k'} j_k j_k' \Delta f_{\alpha,\vec{k}} \Delta f_{\alpha',\vec{k}'}}$ term, representing the occupation-difference-by-momentum-space element sum over two transitions ($\alpha$ and $\alpha'$). The corresponding element in summation over a single one is $\sum_{j_k} j_k \Delta f_{\alpha,\vec{k}}$. Consequently, the definition follows:
\begin{equation}
    \overline{\omega_{\alpha}^2} = \frac{\sum_{j_k} j_k \Delta f_{\alpha,\vec{k}} \omega_{\alpha}^2(\vec{k})}{\sum_{j_k} j_k \Delta f_{\alpha,\vec{k}}}.
\label{eq:mean_omega_alpha}\end{equation}
At this point, a similar approach can also be adopted in the case of two previously discussed quantities, taking:
\begin{align}
    \overline{\tilde{I}_{\alpha,\alpha'}} = \frac{\sum_{j_k} j_k \Delta f_{\alpha,\vec{k}} \tilde{I}_{\alpha,\vec{k},\alpha',\vec{k}'}}{\sum_{j_k} j_k \Delta f_{\alpha,\vec{k}}},\\
    \overline{\int \tilde{\xi}_{\alpha}(x) dx} = \frac{\sum_{j_k} j_k \Delta f_{\alpha,\vec{k}} \int \tilde{\xi}_{\alpha,\vec{k}}(x) dx}{\sum_{j_k} j_k \Delta f_{\alpha,\vec{k}}}.
\end{align}
Now, the missing phases can be recovered as:
\begin{align}
    \phi_{\tilde{I},\alpha,\alpha'} &= \overline{\tilde{I}_{\alpha,\alpha'}} ~/~\left\lvert \overline{\tilde{I}_{\alpha,\alpha'}} \right\rvert,\\
    \phi_{\tilde{\xi},\alpha} &= \overline{\int \tilde{\xi}_{\alpha}(x) dx} ~/~\left\lvert \overline{\int \tilde{\xi}_{\alpha}(x) dx} \right\rvert.   
\end{align}

To sum up, the following equations were obtained in the process of the generalization of the parabolic model to include the $\vec{k}$-dependence of the energies and wavefunctions as well as the $x$-dependence of the material parameters:
\small\begin{multline}
    M_{\alpha,\alpha'} = \delta_{\alpha,\alpha'} \frac{\sum_{j_k} j_k \Delta f_{\alpha,\vec{k}} \omega_{\alpha}^2(\vec{k})}{\sum_{j_k} j_k \Delta f_{\alpha,\vec{k}}} + \phi_{\tilde{I},\alpha,\alpha'} \frac{\Delta_k^2}{2} \times\\
    \times \sqrt{\sum_{j_k} \sum_{j_k'} j_k j_k' \frac{\Delta f_{\alpha,\vec{k}}}{\omega_\alpha(\vec{k})} \frac{\Delta f_{\alpha',\vec{k}'}}{\omega_{\alpha'}(\vec{k}')} \left\lvert \tilde{I}_{\alpha,\vec{k},\alpha',\vec{k}'} \right\rvert^2}
\label{eq:M_matrix_kdp}\end{multline}\normalsize
and
\small\begin{align}
    \beta_j &= \sum_\alpha \mathcal{E}_{\alpha}^j \sqrt{\frac{\Delta_k^2}{2} \sum_{j_k} j_k \frac{\Delta f_{\alpha,\vec{k}}}{\omega_{\alpha}(\vec{k})} \left\lvert \int \tilde{\xi}_{\alpha,\vec{k}}(x) dx \right\rvert^2}~\phi_{\tilde{\xi},\alpha},
\end{align}\normalsize
which are analogous to the Eqs.~(\ref{eq:M_matrix}) and (\ref{eq:beta_j_def}) of the parabolic model, respectively.

\subsubsection{Dipole moment re-scaling}\label{sec:dipole_moment_re-scaling}

Note that while the $S=1$ renormalization mentioned earlier is a well-known and widely accepted way of obtaining the correct dispersion relations and wavefunctions, it leaves an open question of how exactly to compute the inter-subband current elements $\xi_{\alpha,\vec{k}}(x)$. In specific, it is not clear if the original $S$ parameter, the renormalized $S=1$ parameter, or some intermediate effective $S_\text{eff}$ should be used, in each case affecting the $\gamma^{6 \times 6}_i$ in Eq.~(\ref{eq:def_Bloch_band_masses}) and consequently the ${m^\mathcal{B}(x)}$ in Eq.~(\ref{eq:def_tilde_xi_alpha_k}). The latter approach was implemented in Ref.~\cite{pasek2022multisubband}, leading to the correct behavior of the MSP in THz parabola-shaped type-I QW arrays. In this work, the approach based on the dipole matrix moment was adopted.

For a simple effective mass Hamiltonian, it is well known that the transition dipole moment between two states is exactly proportional to the momentum matrix element:
\begin{equation}
    \left\langle \Psi_a \middle| \hat{x} \middle| \Psi_b \right\rangle = \frac{i \left\langle \Psi_a \middle| \hat{p_x} \middle| \Psi_b \right\rangle}{\omega_{a \rightarrow b} m^{*}}
\end{equation}
We could adopt this result in the generalization of the $\xi_\alpha(x)$ quantity, with the following substitution:
\begin{equation}
    \frac{1}{m} \frac{\partial}{\partial x} \rightarrow \omega_\alpha \hat{x},
\end{equation}
which, after including the dependencies on $x$ and $\vec{k}$, leads to
\begin{equation}
    \int \tilde{\xi}_{\alpha,\vec{k}}(x) dx = 2 \omega_{\alpha}(\vec{k}) \int \frac{ \Psi_{n,\vec{k}}(x) \hat{x} {\Psi_{m,\vec{k}}(x)}^{*} }{\sqrt{\epsilon_S(x)}}  dx.
\label{dipole_int_xi}\end{equation}

Let the version of the $\mathcal{L}_{\alpha}$ quantity of Eq.~(\ref{eq:L_alpha}), which was calculated using Eq.~(\ref{dipole_int_xi}) be called $\mathcal{L}^{\hat{x}}_{\alpha}$. Note that $\mathcal{L}^{\hat{x}}_{\alpha}$ does not depend on the different effective masses for different Bloch states. Consequently, it circumvents the effective mass controversy, which arises due to the $S=1$ renormalization mentioned earlier. Thus, it is preferential to use it instead of $\mathcal{L}_{\alpha}$. Unfortunately, there is no relation analogous to Eq.~(\ref{dipole_int_xi}) for the $\tilde{I}_{\alpha,\vec{k},\alpha',\vec{k}'}$, which means the $\hat{x}$ cannot be directly used in the diagonalization of the $M_{\alpha,\alpha'}$ matrix. What can be done however, is the re-scaling of the $\mathcal{D}_{\alpha,\alpha'}$ and the $\mathcal{L}_\alpha$ quantities the following way
\begin{align}
    \tilde{\mathcal{L}}_\alpha &= \mathcal{L}_\alpha {\left\lvert \frac{\mathcal{L}^{\hat{x}}_\alpha}{\mathcal{L}_\alpha} \right\rvert} \nonumber\\
    \tilde{\mathcal{D}}_{\alpha,\alpha'} &= \mathcal{D}_{\alpha,\alpha'} {\left\lvert \frac{\mathcal{L}^{\hat{x}}_\alpha}{\mathcal{L}_\alpha} \right\rvert} {\left\lvert \frac{\mathcal{L}^{\hat{x}}_{\alpha'}}{\mathcal{L}_{\alpha'}} \right\rvert}.
\label{eq:dipole_rescaling_strict}\end{align}
This process leads to results independent of any $S$ renormalization and preserves the correct relative phase orientations of the $\xi$ for any $\alpha$ and $\alpha'$.

Alternatively, a slightly more relaxed assumption can be adopted, that the dipole matrix approach scales to the \ekdp the quantities proportionally, such that
\begin{align}
    \tilde{\mathcal{L}}_\alpha &= \aleph  \mathcal{L}_\alpha {\left\lvert \frac{\mathcal{L}^{\hat{x}}_\alpha}{\mathcal{L}_\alpha} \right\rvert} \nonumber\\
    \tilde{\mathcal{D}}_{\alpha,\alpha'} &= \aleph^2  \mathcal{D}_{\alpha,\alpha'} {\left\lvert \frac{\mathcal{L}^{\hat{x}}_\alpha}{\mathcal{L}_\alpha} \right\rvert} {\left\lvert \frac{\mathcal{L}^{\hat{x}}_{\alpha'}}{\mathcal{L}_{\alpha'}} \right\rvert},
\label{eq:dipole_rescaling_proportional}\end{align}
where $\aleph$ is a scaling constant, identical for all $\alpha$ and $\vec{k}$ and applicable in a broad range of similar heterostructures.

In the single-particle $8$-band Hamiltonian of Sec.~\ref{sec:k.p_single_particle}, the eigenstates are not spin-diagonal due to spin-orbit interaction. Instead, they are Kramer degenerate, with two opposite spin chiralities. In the numerical results, two arbitrary orthogonal linear combinations can be obtained for each Kramer pair, with this combination changing for each $\vec{k}$ value. Since the model treats each subband pair $\alpha$ as a separate element in the matrix, but does sum over $\vec{k}$ for each $\alpha$, the orbital symmetries of the initial and final quantum states for each quantum number $\alpha$ need to be defined consistently as $\vec{k}$ changes. The procedure for this task is described in Appendix~\ref{sec:De-hybridization_of_the_spin_chiralities}. In fact, given the consistently defined chirality, the chirality-preserving transitions are allowed and the cross-chirality transitions are forbidden, thus only a single (allowed) transition needs to be considered between two Kramer pairs, with the doubling of the occupation difference -- similar to the case of spin in the two plasmon models discussed above.

\subsubsection{Simulation results of the \ekdp}

\begin{figure*}[hbt!]
    \centering
    \includegraphics[width=0.34\linewidth]{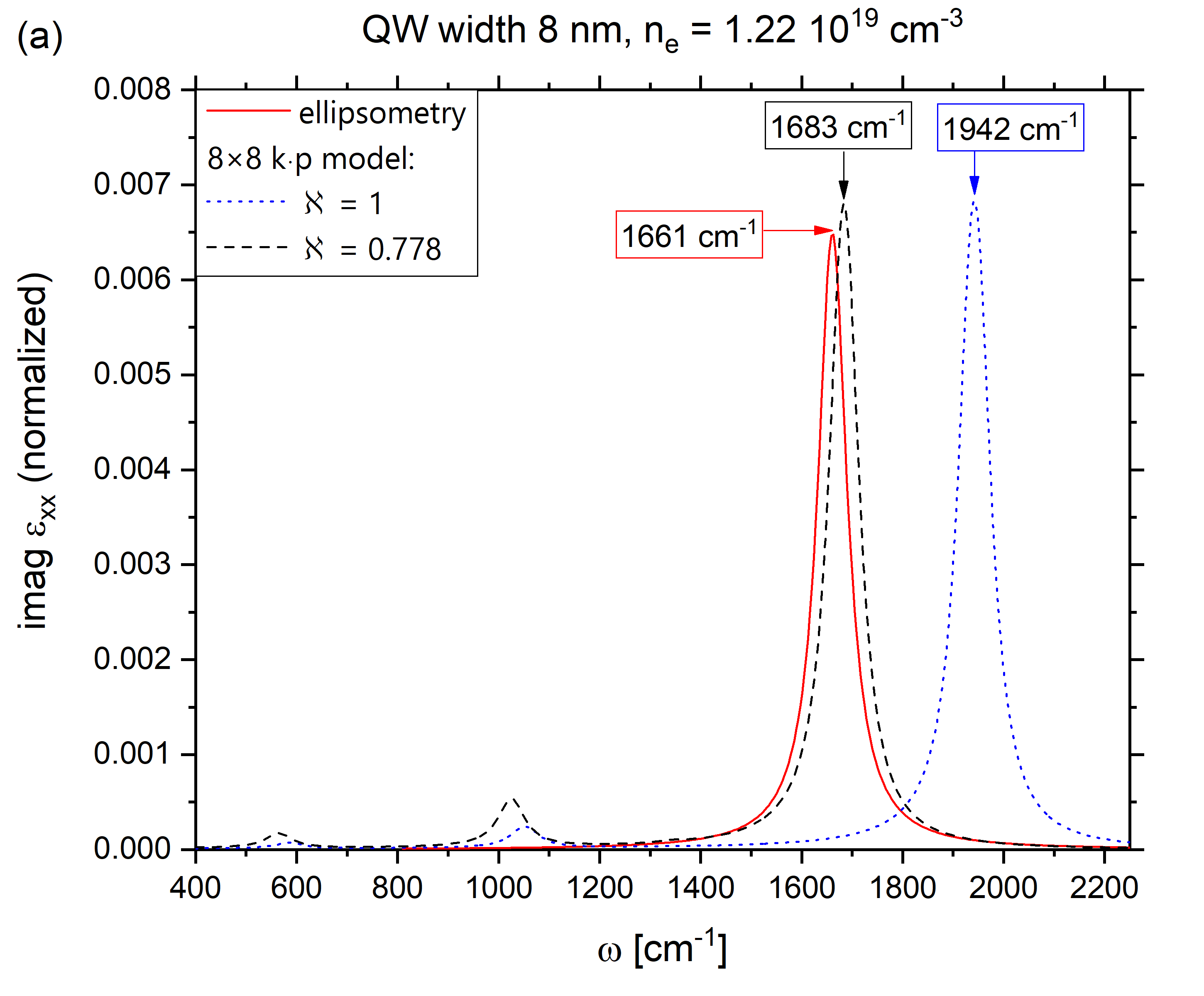}
    \includegraphics[width=0.34\linewidth]{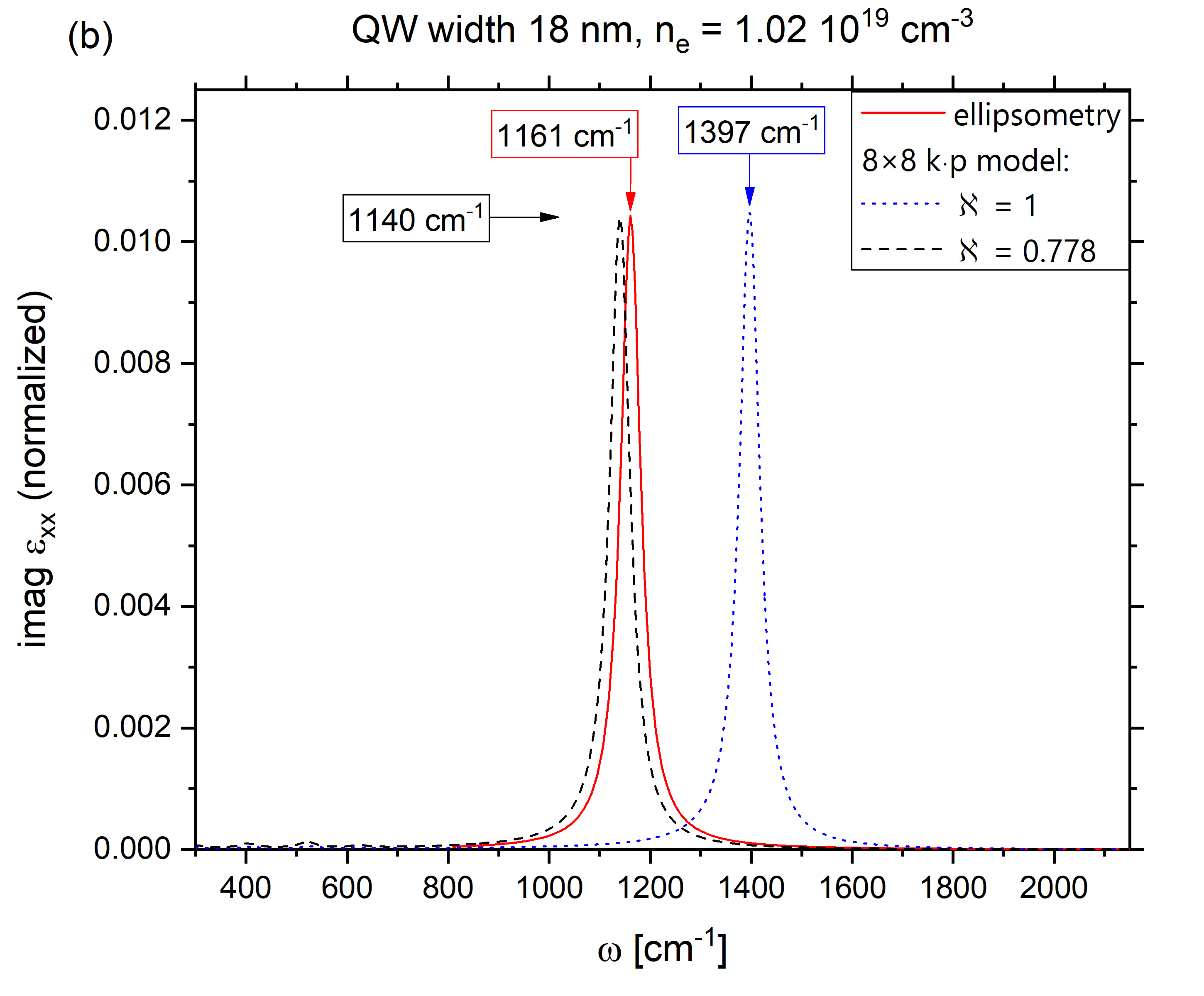}
    \caption{Comparison between the experimentally identified peaks in the imaginary part of the effective dielectric function $\epsilon_{xx}$ (red line) and the corresponding spectra obtained from the \ekdp (black and blue lines). Results shown in (a) are for the $8$~nm QW system, the ones shown in (b) correspond to the $18$~nm QW one. The blue lines show the direct moment re-scaling, following Eq.~(\ref{eq:dipole_rescaling_strict}), while the green line marks the results obtained when allowing for a fitted proportionality factor $\aleph=0.778$ in Eq.~(\ref{eq:dipole_rescaling_proportional}). Note that the spectra were normalized to facilitate direct comparison of the shapes.}
    \label{fig:kdp_vs_experiment}
\end{figure*}

\begin{figure*}[hbt!]
    \centering
    \includegraphics[width=0.35\linewidth]{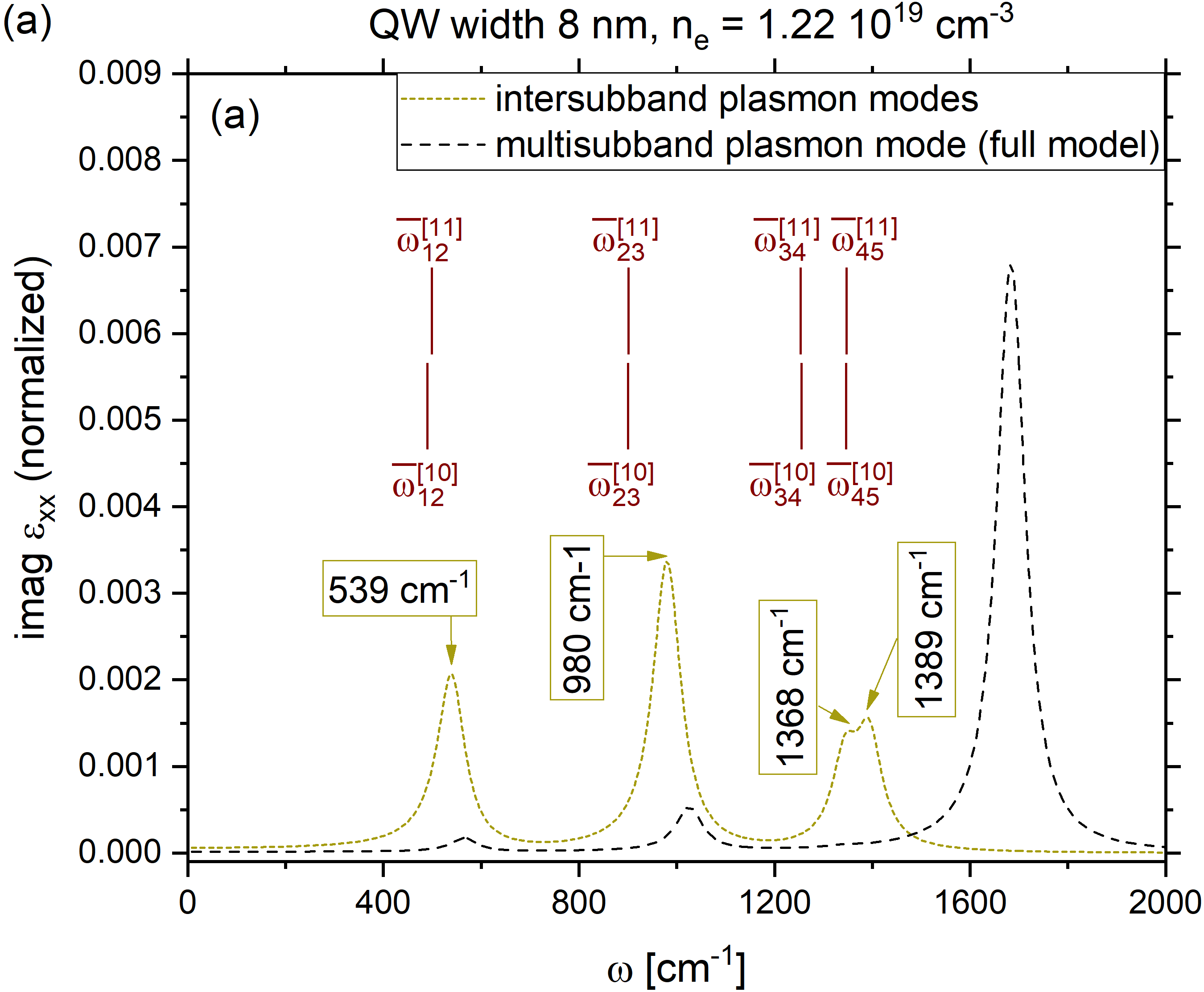}
    \includegraphics[width=0.35\linewidth]{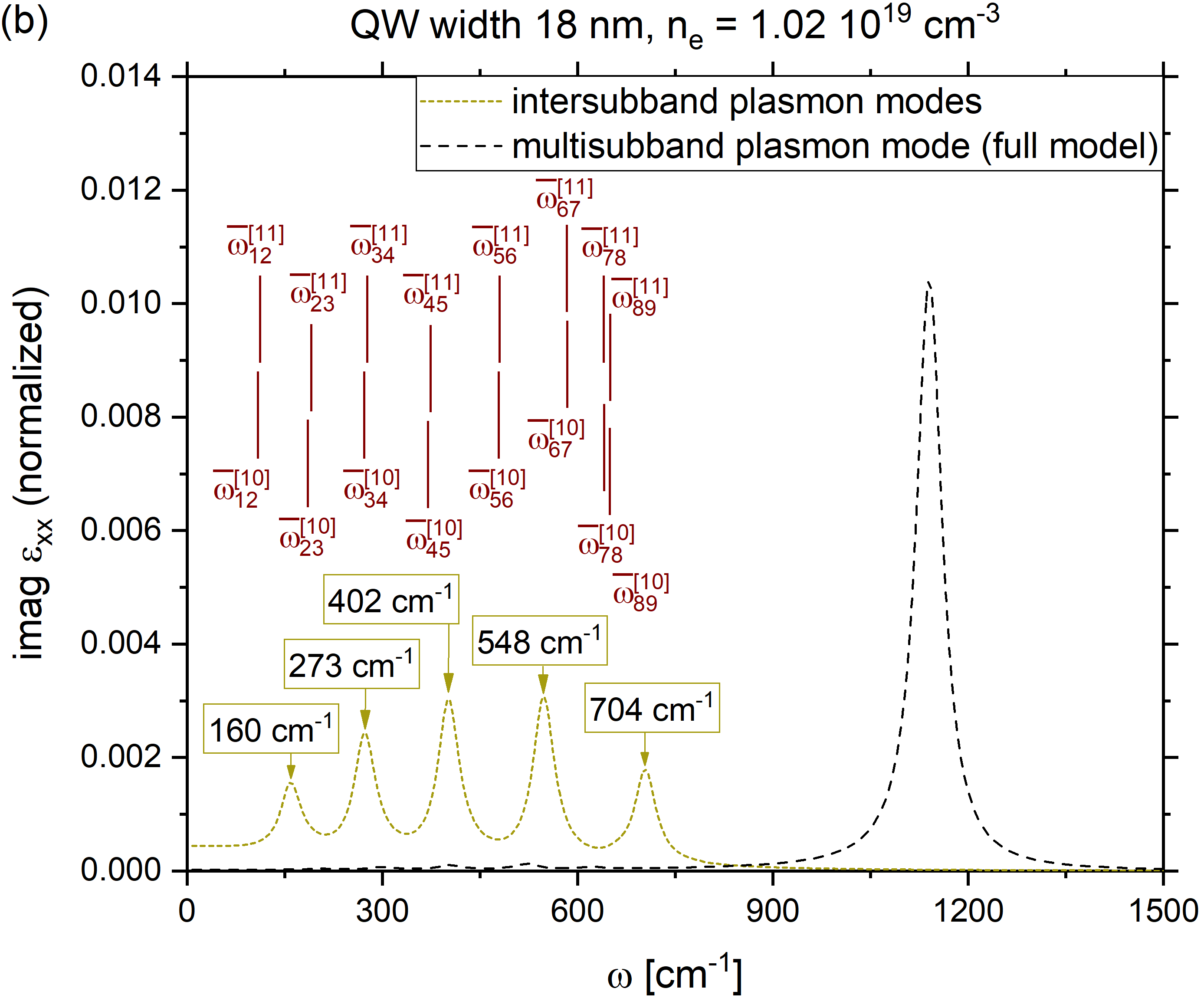}
    \caption{Spectra of the imaginary part of the effective dielectric function $\epsilon_{xx}$ obtained from the \ekdp (a) for the $8$~nm system and (b) for the $18$~nm system. The short-dashed yellow lines shows the spectra for the individual intersubband plasmons. The dashed black line shows the result from the full model, exhibiting a single MSP peak. The vertical brown lines show the averaged intersubband transition energies for individual electrons $\bar{\omega}_\alpha$, corresponding in each case to [10] and [11] orientations of $\vec{k}$.}
    \label{fig:MSP_formation}
\end{figure*}

Figure~\ref{fig:kdp_vs_experiment}(a) shows the \ekdp simulation results for the $8$~nm QW system, while (b) corresponds to the $18$~nm QW system. The green curve represents the simulation performed without accounting for electron-electron interactions. The blue dotted lines mark $\Im(\epsilon_{xx})$ obtained with the direct dipole moment rescaling of Eq.~(\ref{eq:dipole_rescaling_strict}), while the black dashed lines show the results when allowing for the proportional rescaling, as in Eq.~(\ref{eq:dipole_rescaling_proportional}). In the latter case the $\aleph=0.778$ parameter was fitted to minimize the distance between the position of the predicted peaks and the ones obtained experimentally.

It can be noticed that the \ekdp consistently predicts a single main MSP peak, which is qualitatively the same as the ellipsometry result. Moreover, the \ekdp with the direct dipole moment rescaling ($\aleph=1$) predicts the position of these main peaks relatively close to the observed values. Explicitly, the MSP peak is predicted at $1942$\rcm for the $8$~nm QW in comparison to the MSP peak at $1661$\rcm from ellipsometry and the MSP peak is predicted at $1397$\rcm for the $18$~nm QW in comparison to the MSP peak at $1161$\rcm from ellipsometry. These results are already significantly and qualitatively better than the ones obtained with the parabolic or the hybrid model, as can be easily noticed in Fig.~\ref{fig:model_comparison}. Moreover, a simple proportional rescaling of $\aleph=0.778$ allows to obtain results in qualitative agreement with the experiment.
As shown in Figs.~\ref{fig:kdp_vs_experiment}(a) and (b), the difference in the central frequency between the dashed black line peaks and solid red line peaks are only $22$\rcm and $21$\rcm for the $8$~nm QW and $18$~nm QW systems, respectively.

Figure~\ref{fig:MSP_formation} investigates in detail how the single MSP peak is formed in the \ekdp. In this figure, the results for the three stages of the plasmon mode formation are shown. The brown vertical lines show the mean IST frequency $\overline{\omega_\alpha}$ of the transitions, with the initial and final subband quantum number shown, as well as the orientation of the Brillouin zone subspace $d \in \lbrace [10], [11] \rbrace$. The short-dashed yellow lines in the figure correspond to the formation of the purely ISP modes, when the single-particle oscillators of an individual $n_I \rightarrow n_F$ subband pair are coupled, but there is no coupling across the different subband pairs. Mathematically, this is equivalent to putting all the off-diagonal terms in Eq.~(\ref{eq:M_matrix_kdp}) to zero, with the exception of the ones corresponding to the same $n_I \rightarrow n_F$. Finally, the dashed black line for the final \ekdp result is added for comparison. It can be noticed that the ISP spectra coincide with the IST frequencies, as it was the case for the parabolic model (compare Fig.~\ref{fig:parabolic_analysis}) and in the case of the hybrid model (see Fig.~\ref{fig:hybrid_analysis}). In the other models however, this was the case in the final result, when the coupling between the individual ISP was mathematically allowed, yet it was weak enough not to lead to the formation of a single dominating MSP, which is opposite to the \ekdp. This difference comes from the coincidence of two facts discussed below.

Firstly, the values of the IST frequencies $\overline{\omega_\alpha}$ are much lower in the case of \ekdp in comparison with the two other models, for both the $8$~nm QW and the $18$~nm QW systems.
To be specific, for the former system, the central frequencies of the $\overline{\omega_\alpha}$ peaks lie within the $2930-4343$\rcm interval for the parabolic model, in a similar $2795-4200$\rcm interval for the hybrid model, but in the $490-1347$\rcm interval in case of the \ekdp. It particularly stands out that even the minimal $\overline{\omega_\alpha}$ in the results of the first two models is already bigger than the experimentally observed MSP position of $1661$\rcm, while the maximal $\overline{\omega_\alpha}$ value for the \ekdp lies below the frequency of the experimentally detected peak. This is especially significant, as the collective plasma effect, in general, blueshifts the transition frequency in comparison to the single-particle ones. Thus, it is clear that the single-band Hamiltonian strongly overestimates the $\overline{\omega_\alpha}$ frequencies. In the case of the $18$~nm QW system, the $\omega_\alpha^0$ lie in the $802-2326$\rcm range in the parabolic model, while $\overline{\omega_\alpha}$ lie in the $778-2310$\rcm range in the hybrid model, but in the $108-650$\rcm range in the \ekdp. In this case, some of the IST frequencies are found both below the experimentally observed peak at $1161$\rcm for the models using the single-band Hamiltonian, but still most of the single-particle frequencies lie above this value. On the other hand, the multi-band Hamiltonian places the frequencies firmly below the MSP peak detected by ellipsometry.

Secondly, the differences between the $\overline{\omega_\alpha}$ frequencies of different $n_I \rightarrow n_F$ transitions are much smaller in the case of \ekdp than the two other models. Explicitly, in the case of the $8$~nm QW system, the frequency separations of sequential subband pairs lie in the $91-401$\rcm range in the \ekdp, but are $1413$\rcm for the parabolic model and $1405$\rcm for the hybrid model. In the case of the $18$~nm QW system, the corresponding ranges are $9-105$\rcm for \ekdp, $467-536$\rcm for the parabolic model and $478-529$\rcm for the hybrid model. As the $\overline{\omega_\alpha}$ values lie closer to one another, then the plasma shifts are relatively stronger in respect to their separations, leading to an increase in coupling between the modes. In the strong coupling regime, a single dominating MSP mode is formed, which corresponds to the constructive interference of all the constituent oscillations and that "absorbs" nearly all the light-matter coupling intensity from the other modes.
In fact, the other MSP modes can also be found in the \ekdp $\Im(\epsilon)$ spectra (black dashed line, blue dotted line) of the $8$~nm QW system in Figs.~\ref{fig:kdp_vs_experiment}(a) and \ref{fig:MSP_formation}(a), at around $~600$\rcm and $~1000$\rcm. These peaks are strongly diminished as a result of negative interference between the constituent oscillations, which is due to the orthogonality requirement (i.e. there can be only one fully constructive MSP mode). In the case of the $18$~nm QW system, these nodes can be found in the data, but are not visible in the scale of Figs.~\ref{fig:kdp_vs_experiment}(b) and \ref{fig:MSP_formation}(b).
The both discussed facts (difference in scales of $\overline{\omega_\alpha}$ values and their separations) come from the significantly different single-particle dispersion relation in the \ekdp, in comparison to the single-band Hamiltonian, as was explained before.

\section{Conclusion}\label{sec:conclusion}
In this work, the collective optical response of heavily doped InAs / GaSb broken gap QW has been studied experimentally and theoretically.
The plasmon modes of the samples were studied experimentally using ellipsometric spectroscopy, with the experimental data interpreted with the Lorentz model, leading to observation of single dominating Multi-subband Plasmon Mode (MSP), as described in Sec.~\ref{sec:spectroscopic_ellipsometry}.
On the theoretical side, the limitations of the semiclassical parabolic model in describing the InAs/GaSb QW were highlighted in Sec.~\ref{sec:parabolic_model}.
This model relies on the single-band description of the single-particle dispersion relation and the corresponding wavefunctions, described in Sec.~\ref{sec:single-band_single_particle}, and does not account for non-parabolicity and the conduction-valence mixing of the bulk Bloch bands, which are key characteristics of this type of QW.
In explicit, no strong coupling was found between the ISPs corresponding to individual ISTs, which lead to the MSP not being formed, in stark qualitative contrast to the experimental data.
Next, it was shown in Sec.~\ref{sec:hybrid_model} that the results were only marginally improved in the hybrid model, but still qualitatively different from the expected outcome.
The latter model considers the subband non-parabolicity resulting from the spatial inhomogeneity of the bulk parameters, which is due to the heterostructure being composed of two materials, but still neglecting the inter-band mixing.
Finally, the semiclassical approach was generalized adopting the $8$-band formalism at the single-particle level, as given in Sec.~\ref{sec:k.p_single_particle}.
This generalization allows for the inclusion of both dispersion relation non-parabolicity and inter-band mixing and leads to dependence of both the transition frequencies and the single-particle wavefunctions on the in-plane wavevector $\vec{k}$.
In Sec.~\ref{sec:ekdp_model}, it was demonstrated that the \ekdp is the only one able to qualitatively explain the experimental observation.
Furthermore, with a single rescaling parameter $\aleph$, a perfect quantitative agreement in the frequency of the MSP can be achieved between the \ekdp and the ellipsometry results.
Additionally, it was proven that no rescaling of the single-band effective mass in the hybrid model can lead to even qualitatively correct results for both samples, see Appendix~\ref{sec:single-band_arbitrary_mass}.

The discrepancies in the predictions between the former two models and the \ekdp come from the different way it interprets the curvature of the conduction bulk bands of the both materials. The single-band Hamiltonian comes from the perturbation theory, with the conduction band effective mass value being an effect of the remote bands contributions. In the \ekdp single-particle Hamiltonian, there is the group of $8$ Bloch bands incorporated directly in the basis set, with the rest of the bulk bands treated perturbatively. In general, the former (simpler) approach works well for the type-I structures, where the bulk conduction and valence bands of the both materials are well separated energetically. It fails however in the case of a heterostructure with a broken gap, where the inter-band mixing is very strong due to quantum size effect. In fact, this should not be surprising, as it is a widely known fact that the perturbative approach stops working as the levels involved approach near degeneracy. The methodology presented in this work is not particular to the systems presented but can be adapted to work with type-III QWs in general.

\section{Discussion}\label{sec:discussion}

It may be useful to discuss the description of band non-parabolicity adopted in this work in comparison to previous approaches. In general, the inclusion of non-parabolic effects in the quantum model follows the approach presented in \cite{askenazi2014ultra-strong}, where the plasma shift frequency is obtained as
\begin{equation}
\omega_{P,n}^2(k_{||}) dk_{||} = \frac{f_n e^2 \left(dN_n(k_{||}) - dN_{n+1}(k_{||})\right)}{m^{*}(k_{||}) \epsilon_0 \epsilon_{\infty} L_{\text{eff},n}}
\label{eq:quantum_non-parabolicity}\end{equation}
—which corresponds to Eq.~(2) of the cited work. The non-parabolicity enters this equation in two ways: through the dependence of the energy difference between subbands and of the effective mass on $k_{||}$. This is analogous to how non-parabolicity was originally incorporated into the semiclassical model; see Eq.~(3) of Ref.~\cite{alpeggiani2014semiclassical}.

It is worth noting that in Eq.~(\ref{eq:quantum_non-parabolicity}), the relevant quantities depend only on the magnitude of the in-plane momentum, not its orientation. Furthermore, the mixing between conduction and valence bands is omitted—both in the form of the wavefunction (and, consequently, the calculation of intersubband current elements $\xi_{n\rightarrow n+1}$) and in the use of scalar effective masses $m^{*}(k_{||})$. In fact, the tensor effective masses of different bands (cb, hh, lh, so) differ from the simple bulk conduction band effective mass. Additionally, the discontinuity of mass parameters at the interface of the two materials does not appear to be explicitly considered in Eq.~(\ref{eq:quantum_non-parabolicity}), nor is the dependence of the effective mass on the subband index $n$ \footnote{However, this may result from the simplified notation of Eq.~(\ref{eq:quantum_non-parabolicity}), as the position dependence of $m^*$ is explicitly accounted for in \cite{warburton1996intersubband}, which is cited in Ref.~\cite{askenazi2014ultra-strong} as the source for incorporating non-parabolicity.}.
Keeping in mind the distinctions between the quantum and semiclassical models, this approach is roughly equivalent to the hybrid model discussed in this work. All the aforementioned simplifications are justified provided that the nanostructure under investigation is a simple type-I square QW. In such systems, the impact of conduction–valence mixing, as well as well–barrier hybridization, is negligible due to the large band gap separation. However, the applicability of this model is necessarily limited, excluding type-III QWs, as demonstrated in the present paper.

Another example where the approach of Eq.~(\ref{eq:quantum_non-parabolicity}) leads to unsatisfactory results is in QWs with a parabolic potential profile, as shown in \cite{pasek2022multisubband}. In that case, the depolarization shift nearly vanishes due to the Kohn theorem \cite{kohn1961cyclotron} (aka harmonic potential theorem). Since this shift should vanish analytically for a perfect harmonic oscillator, small higher-order terms in the $\kdp$ Hamiltonian dominate the plasmon shift dynamics and cannot be neglected, unlike in simple square type-I QWs. Taking into account both Ref.~\cite{pasek2022multisubband} and the present work, it can be concluded that the \ekdp successfully captures the impact of band non-parabolicity on the frequency of MSP modes in QWs beyond the regime where Eq.~(\ref{eq:quantum_non-parabolicity}) applies.

A recent example of accounting for the impact of band non-parabolicity on the plasmonics of doped semiconductor systems can be found in a fairly recent PhD thesis \cite{haky2022semiconductor}. In the cited dissertation, the $8\times8~\vec{k}\cdot\vec{p}$ single-particle description of electrons was obtained using the \textit{nextnano} software, as in the present work. However, a $3$-band approach, similar to Eq.~(\ref{eq:quantum_non-parabolicity}), was adopted to construct the plasmon model, rather than taking advantage of the $k_{||}$-dependent wavefunctions and the explicit form of the matrix Hamiltonian.

As a general remark, it is important to note that the non-parabolicity of the energy dispersion arises fundamentally from band mixing. The perturbative approach attempts to describe this resulting effect without explicitly accounting for its underlying cause. Such an indirect treatment inevitably limits the range of applicability, thereby highlighting the need for a more general and comprehensive approach.

\section{Outlook and perspectives}\label{sec:Outlook_and_perspectives}

Regarding further theoretical developments, at least two directions are possible. Based on the present results, a semiclassical description of more advanced type-III systems—beyond a single QW—can be pursued. In such systems, MSPs may be fine-tuned by external parameters such as electric field or temperature, taking advantage of the underlying single-particle dynamics, which are qualitatively different from those in type-I structures. Explicit inclusion of the $8\times8~\vec{k}\cdot\vec{p}$ single-particle Hamiltonian in the quantum model is also feasible. Admittedly, the quantum model is more powerful in the broader context, as it enables the treatment of individual plasmonic quasiparticles and naturally integrates with systems including cavity modes, phonon coupling, nonlinear effects, and other phenomena. There is no fundamental obstacle to constructing a multi-band quantum model; the challenges are primarily computational—such as diagonalizing very large matrices—but can be overcome with the aid of high-performance computing (HPC).

\section{Acknowledgments}\label{sec:Acknowledgments}

This work was supported by the OCP Foundation through the Chair "Multiphysics and HPC," led by Mohammed VI Polytechnic University. Additional support was provided by the UM6P starting package 163DOPR02-2. We gratefully acknowledge the use of the High-Performance Computing (HPC) facilities at UM6P—Toubkal. This work was partially funded by the French Occitanie Region (SEA - ESR-PREMAT-238), the French Agence Nationale pour la Recherche (EquipEx EXTRA, ANR 11-EQPX-0016, Equipex HYBAT, ANR-21-ESRE-0026, NanoElastir, ASTRID 2020–2023), and the I-Site MUSE (ENVIRODISORDERS).

\appendix
\section{Material parameters used in the simulation}\label{sec:material_parameters}

\begin{table}[!ht]
    \centering
    \begin{tabular}{|c|c|c|c|c|}
        \hline
        parameter & InAs value & GaSb value & units & source \\
        \hline
        $E_G^{0\text{K}}$ & $0.417$ & $0.812$ & eV & \cite{vurgaftman2001band} \\
        \hline
        $\alpha$ & $0.276$ & $0.417$ & meV/K & \cite{vurgaftman2001band} \\
        \hline
        $\beta$ & $93.0$ &  $140.0$ & K & \cite{vurgaftman2001band} \\
        \hline
        $E_P$ & $21.5$ & $27.0$ & eV & \cite{vurgaftman2001band} \\
        \hline
        $\Delta_{SO}$ & $0.39$ & $0.76$ & eV & \cite{vurgaftman2001band} \\
        \hline
        $m_e^{*}$ & $0.026$ & $0.039$ & $m_0$ & \cite{vurgaftman2001band} \\
        \hline
        $E_\text{VBO}$ & $-0.59$ & $-0.03$ & eV & \cite{vurgaftman2001band} \\
        \hline
        $\gamma^{6\times6}_1$ & $20.0$ & $13.4$ & 1 & \cite{vurgaftman2001band} \\
        \hline
        $\gamma^{6\times6}_2$ & $8.5$ & $4.7$ & 1 & \cite{vurgaftman2001band} \\
        \hline
        $\gamma^{6\times6}_3$ & $9.2$ & $6.0$ & 1 & \cite{vurgaftman2001band} \\
        \hline
        $\epsilon_\text{lf}$ & $15.15$ & $15.70$ & $\epsilon_0$ & \cite{NSMarchive} \\
        \hline
        $\epsilon_\text{hf}$ & $12.30$ & $14.40$ & $\epsilon_0$ & \cite{NSMarchive} \\
        \hline
    \end{tabular}
    \caption{Material parameters used in the simulation of the single-particle system and the plasmon modes.}
    \label{tab:material_parameters}
\end{table}

\begin{table}[!ht]
    \centering
    \begin{tabular}{|c|c|c|c|c|}
        \hline
        parameter & InAs value & GaSb value & units \\
        \hline
        $E_G^{300\text{K}}$ & $0.354$ & $0.727$ & eV \\
        \hline
        $\tilde{E}_P^{300\text{K}}$ & $16.06$ & $21.584$ & eV \\
        \hline
        $S$ & $1$ & $1$ & $\frac{\hbar^2}{2 m_0}$ \\
        \hline
        $B$ & $0$ & $0$ & $\frac{\hbar^2}{2 m_0}$ \\
        \hline
        $L^{300\text{K}}$ & $-9.605$ & $-3.498$ & $\frac{\hbar^2}{2 m_0}$ \\
        \hline
        $M$ & $-4$ & $-5$ & $\frac{\hbar^2}{2 m_0}$ \\
        \hline
        $N^{300\text{K}}$ & $-9.805$ & $-6.298$ & $\frac{\hbar^2}{2 m_0}$ \\
        \hline
        $\kappa^{300\text{K}}$ & $-0.033$ & $-0.950$ & 1 \\
        \hline
    \end{tabular}
    \caption{Room temperature parameters of the \ekdp in the case of the $S=1$ renormalization.}
    \label{tab:rescaled_DKK_parameters}
\end{table}

The material parameters used in this work are shown in Table~\ref{tab:material_parameters}. The \ekdp DKK parameters can be obtained directly from the unmodified Luttinger parameters $\gamma^{6\times6}_i$, the energy gap $E_G^T$ and the inter-band Kane energy $E_P$. It must be taken into consideration that the energy gap changes with the temperature $T$, as described by the well-known Varshni formula
\begin{equation}
    E_G^T = E_G^{0\text{K}} - \frac{\alpha T^2}{T+\beta}
\label{eq:Varshni_formula}\end{equation}
Additionally, in the case of the $S=1$ renormalization, the $E_P$ needs to be rescaled as follows.
\begin{equation}
    \tilde{E}_P^{T} = \left(\frac{m_0}{m_e^{*}}-1\right) \frac{E_G^T (E_G^T + \Delta_{SO})}{E_G^T + \frac{2}{3} \Delta_{SO}}
\end{equation}

This rescaling makes the Kane energy $\tilde{E}_P^{T}$ dependent on temperature via Eq.~(\ref{eq:Varshni_formula}), and consequently makes the DKK parameters also vary with temperature. Their values, which were used in the \ekdp simulation for $300$~K and $S=1$, are given in Table~\ref{tab:rescaled_DKK_parameters}. The details on the relation between the modified $\gamma^{8\times8}_i$ and unmodified $\gamma^{6\times6}_i$ Luttinger parameters, on the mathematical impact of the $S=1$ rescaling, and on the temperature dependence can be found in Ref.~\cite{birner2011modeling}.

\section{De-hybridization of the spin chiralities}\label{sec:De-hybridization_of_the_spin_chiralities}

Due to the Kramer degeneracy of each level pair corresponding to the same orbital, but opposed spin chiralities, the output wavefunctions are in general their mixed linear combinations in the \textit{nextnano++} output. This would not matter in the case of the model where each transition between the levels $n$ and $m$ at each $\vec{k}$ was treated as a separate oscillator $\beta = (n,m,\vec{k})$. This, however, would lead to eigenproblem of a very large matrix $M_{\beta,\beta'}$. Instead, in the \ekdp, each $n \rightarrow m$ inter-subband transition is considered as a single oscillator for all $\vec{k}$ along $[10]$ and a single oscillator for all $\vec{k}$ along $[11]$. In this case, one needs to make sure that the momentum space sum over states is consistent for a given $\alpha$, that is, without (I) random changes in the spin chirality compositions of the eigenfunctions of the same $\alpha$ and (II) phase discontinuities between the eigenstates of the same $\alpha$.

The following procedure allows to de-hybridize the degenerate eigenfunctions and sort them into two groups of opposing spin chirality. Let $\Psi_{a,\vec{k}}$ and $\Psi_{b,\vec{k}}$ denote the mixed wavefunctions. The elimination factor $\eta_\mathcal{B}$ for the Bloch band $\mathcal{B}$ can be defined as
\begin{align}
    \eta_{\mathcal{B},\vec{k}} = \frac{\left\langle \Psi_{a,\vec{k}}^\mathcal{B} \middle| \Psi_{a,\vec{k}}^\mathcal{B} \right\rangle}{\left\langle \Psi_{a,\vec{k}}^\mathcal{B} \middle| \Psi_{b,\vec{k}}^\mathcal{B} \right\rangle}.
\end{align}
Next, the wavefunction with $+$ spin chirality will be defined as this linear combination of the $\Psi_a$ and $\Psi_b$, which minimizes the $\left\lvert S\downarrow_x \right\rangle$ component and the wavefunction with $-$ spin chirality as the one with the minimal $\left\lvert S\uparrow_x \right\rangle$ component. They can be obtained by
\begin{align}
        \Psi_{+,\vec{k}} &= C_{+,\vec{k}} \left( \left| \Psi_{a,\vec{k}}  \right\rangle - \eta_{\left\lvert S\downarrow_x \right\rangle,\vec{k}} \left| \Psi_{b,\vec{k}}  \right\rangle \right) \nonumber\\
        \Psi_{-,\vec{k}} &= C_{-,\vec{k}} \left( \left| \Psi_{a,\vec{k}}  \right\rangle - \eta_{\left\lvert S\uparrow_x \right\rangle,\vec{k}} \left| \Psi_{b,\vec{k}}  \right\rangle \right),
\end{align}
where $C_{+,\vec{k}}$ and $C_{-,\vec{k}}$ are the normalization constants. Thanks to this procedure, a given $\alpha$ transition in each case (i.e. for all $|\vec{k}|$ along a given $n$, $m$, and $\vec{k}$ direction) refers to the transition between two levels of defined chiralities and is consistently a chirality-aligned ($+ \rightarrow +$ or $- \rightarrow -$) or a cross-chirality transition ($+ \rightarrow -$ or $- \rightarrow +$).

Furthermore, the wavefunctions corresponding to the same level can experience a random general phase rotation. The general phase for all relevant $|\vec{k}|$ can be made consistent by finding, for each $|\vec{k}|$, the phase rotation factor $\exp\left(i \phi_{\pm,\vec{k}} \right)$, which minimizes the imaginary part of the dominating conduction Bloch component, that is the quantity:
\begin{equation}
    \int \Im\left( \exp\left(i \phi_{+,\vec{k}} \right) \Psi_{+,\vec{k}}^{\left\lvert S\uparrow_x \right\rangle}(x) \right)^2 dx
\end{equation} 
for the $+$ chirality and the quantity:
\begin{equation}
    \int \Im\left( \exp\left(i \phi_{-,\vec{k}} \right) \Psi_{-,\vec{k}}^{\left\lvert S\downarrow_x \right\rangle}(x) \right)^2 dx
\end{equation}
in the case of the $-$ chirality. Finally, the general phase of the eigenfunctions can be re-defined as
\begin{equation}
    \tilde{\Psi}_{\pm,\vec{k}} = \exp\left(i \phi_{\pm,\vec{k}} \right) \left| \Psi_{\pm,\vec{k}} \right\rangle.
\end{equation}
This makes the $\tilde{\Psi}_{\pm,\vec{k}}$ maximally aligned with the real axis, but it still allows of the random $\pi$-swaps of the general phase. Finally, the latter can be controlled by choosing a single $\vec{k_b}$-mesh point as a reference basis and then investigating the sign of the $\left\langle \tilde{\Psi}_{\pm,\vec{k_b}} \middle| \tilde{\Psi}_{\pm,\vec{k}} \right\rangle$ overlap factor.

\section{The hybrid model with arbitrary effective mass}\label{sec:single-band_arbitrary_mass}

\begin{figure*}[hbt!]
    \centering
    \includegraphics[width=0.35\linewidth]{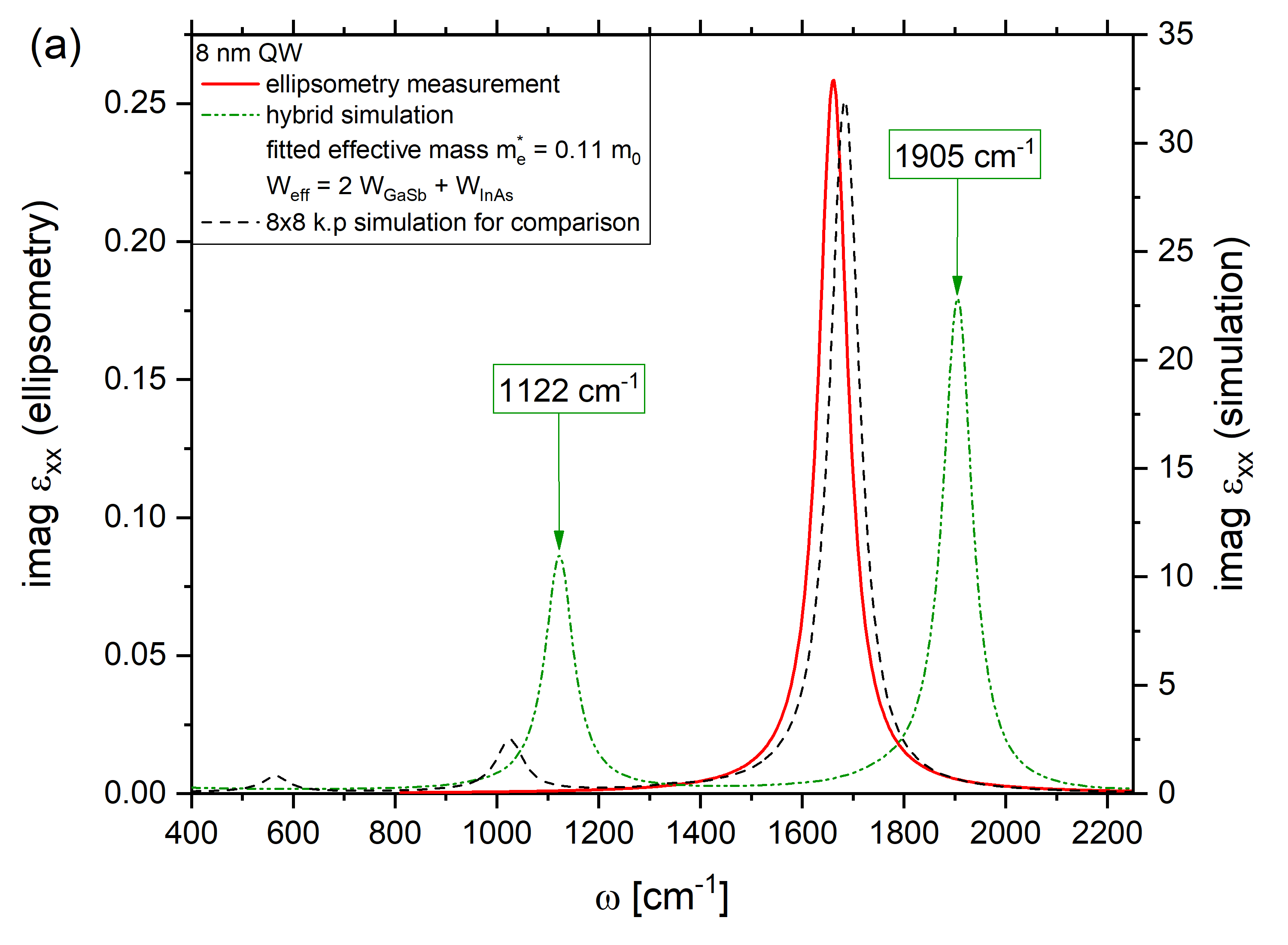}
    \includegraphics[width=0.35\linewidth]{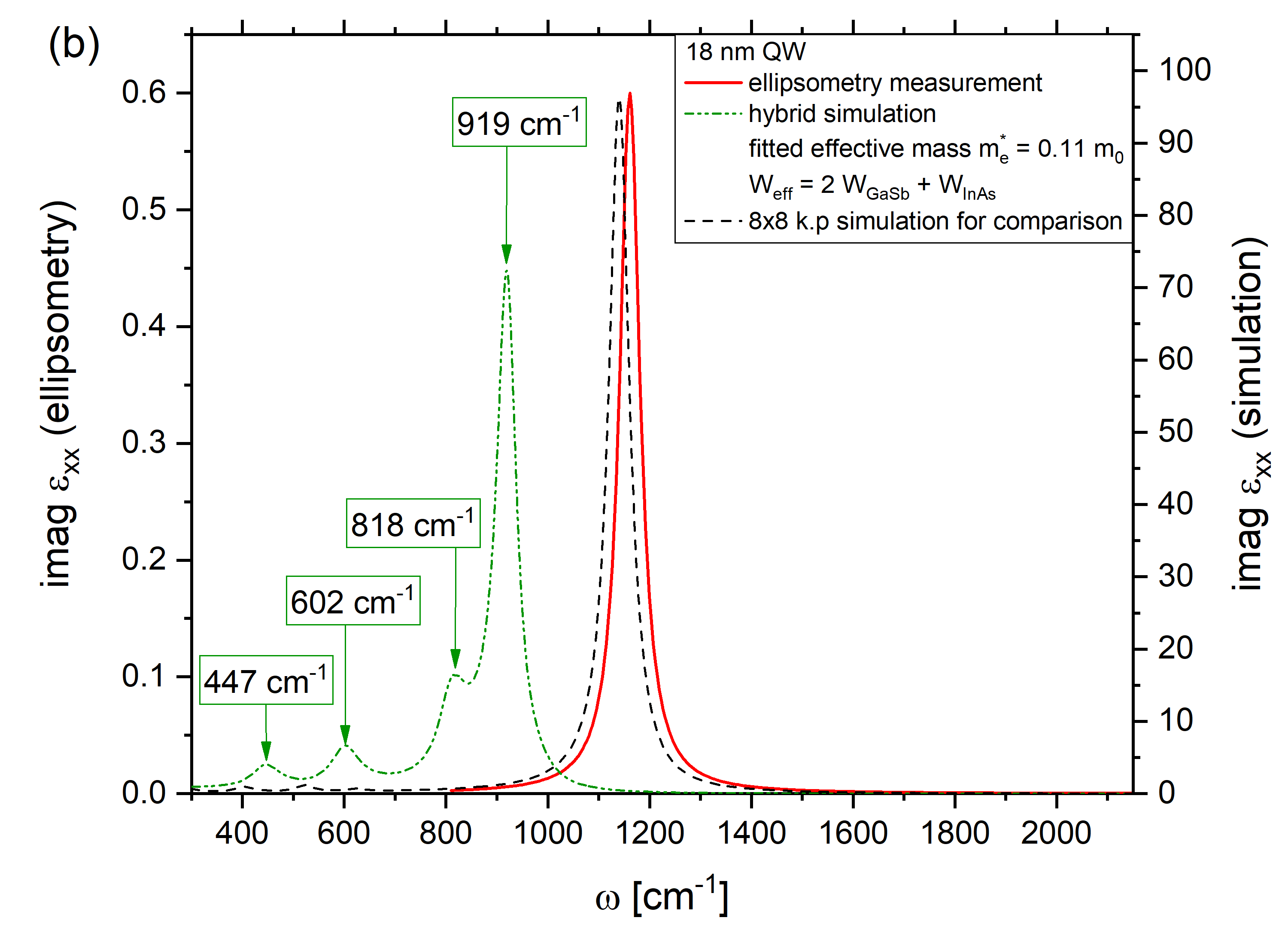}
    \includegraphics[width=0.35\linewidth]{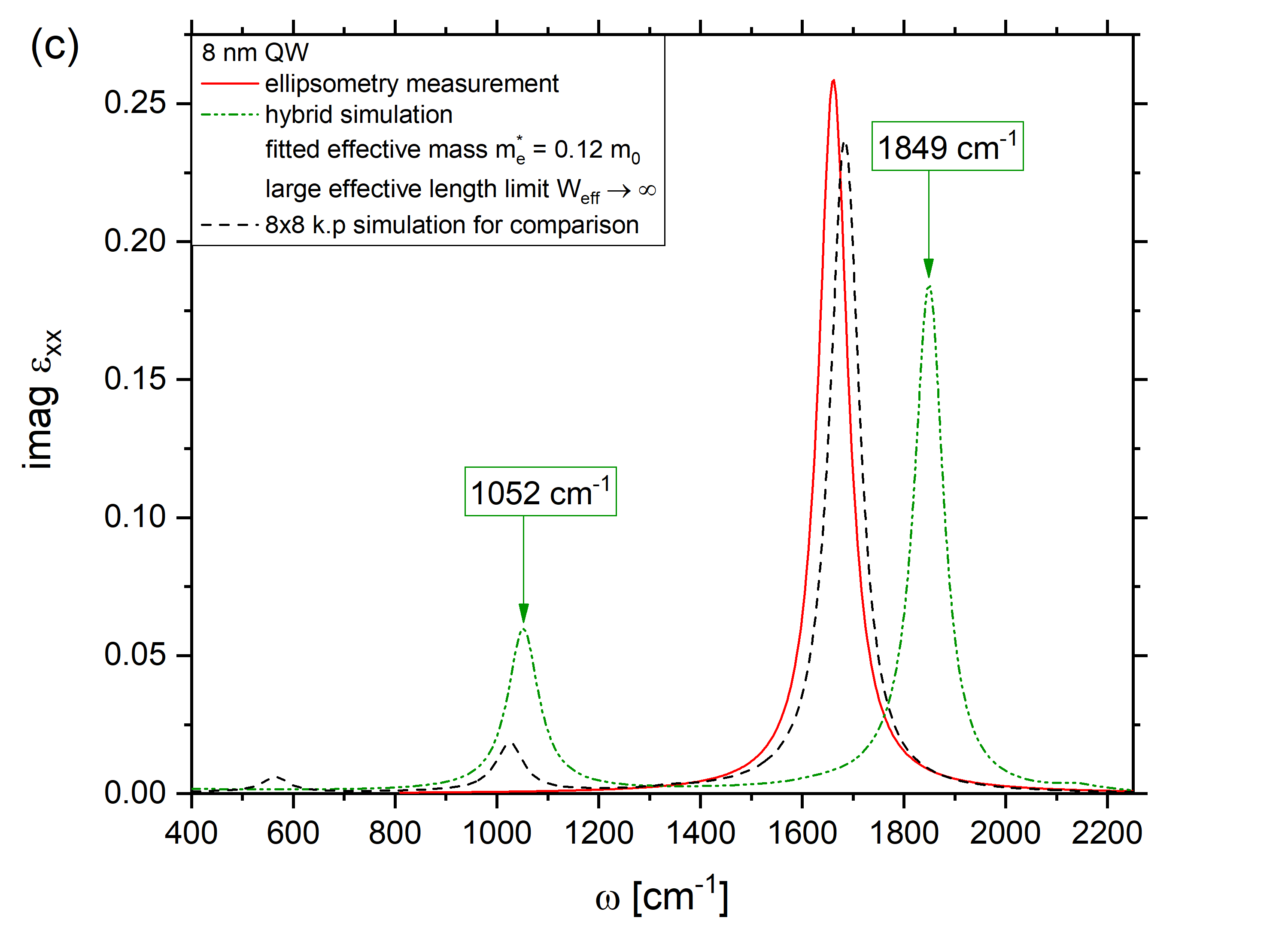}
    \includegraphics[width=0.35\linewidth]{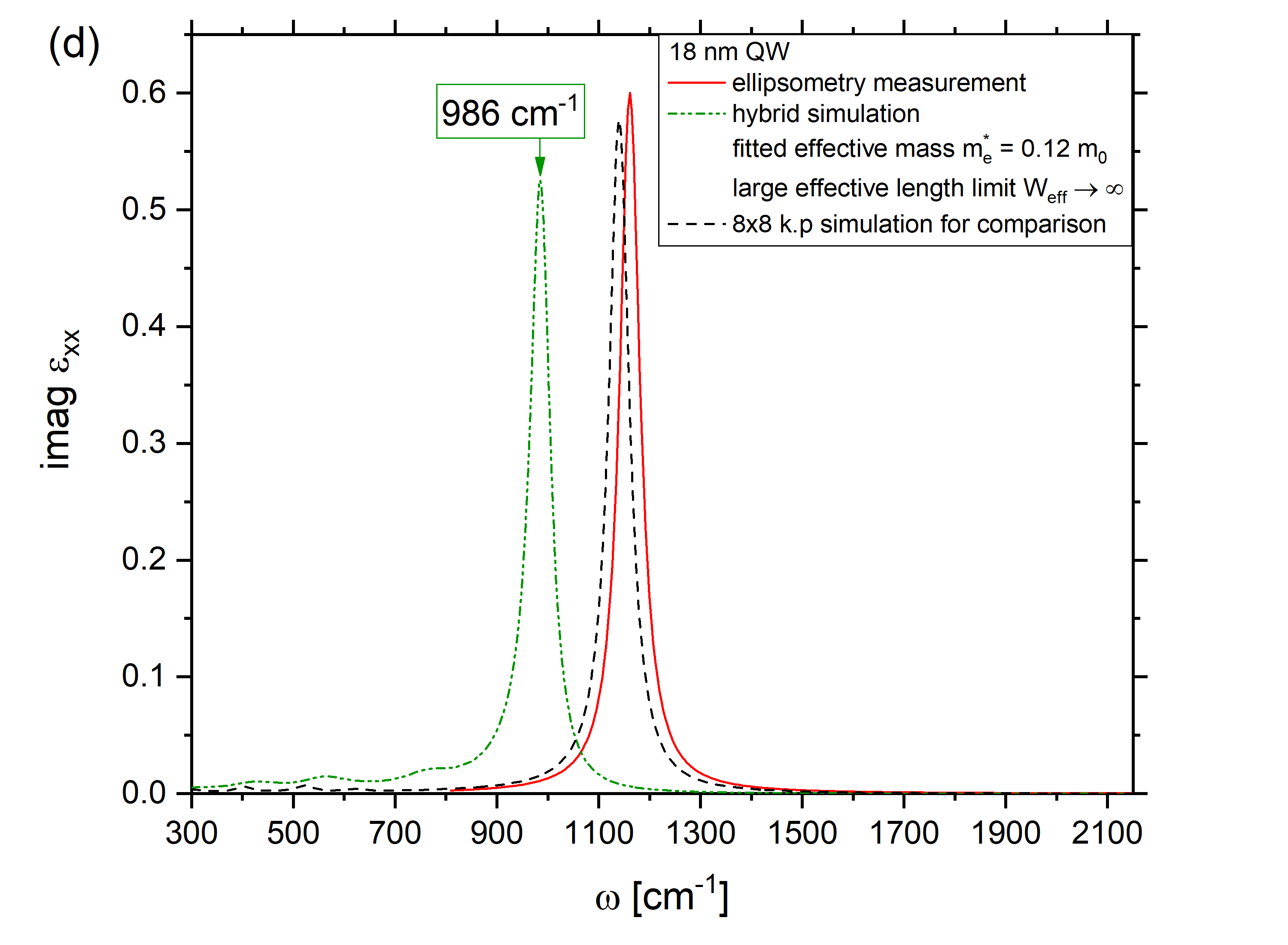}
    \caption{Fitting of the hybrid model to the experimental results, with the effective mass $m_\text{InAs}^*$ as the free parameter. Imaginary parts of the dielectric function are shown with the green dash-dot-dotted lines. (a), (c): for the QW of $8$~nm width; (b), (d): for the QW of $18$~nm width. The $W_\text{eff}$ in the effective medium approximation corresponds to the exact size of the GaSb/InAs nanosystem in (a), (b), while the $W_\text{eff}\rightarrow\infty$ limit is considered in (c) and (d). The black dotted lines in (a), (b) and (c) show the shape of the spectrum for the \ekdp and the red solid lines show the experimental ellipsometry results, for comparison.}
    \label{fig:single-band-fit}
\end{figure*}

The very large overestimation of the MSP resonant frequency in the case of the parabolic and the hybrid models can be attributed, in the most part, to an incorrect description of the singe-particle dispersion relation given by the single-band model based on the bulk parameters. In the case of bulk materials, the large values of energy gap $E_G$ of both materials keep the conduction bands well separated from the valence bands. However, in the case of type-III (broken gap) heterostructure, there is an energy range where the conduction bands of the InAs well overlap with the valence bands of the GaSb barrier, which leads to hybridization of different Bloch bands and formation of mixed conduction-valence eigenstates. In the case of the particular nanosystems under investigation, as the result of significant n-type doping, the Fermi level lies above the hybridization regime. However, the dispersion relation of the conduction bands is still strongly affected by the broken gap. As an effect, the bulk effective mass parameters $m^*$ for the conduction bands do not capture correctly the underlying physics of the broken gap nanostructure. One may hypothesize that different effective $m^*$ values can encapsulate this dynamic correctly, allowing to obtain the experimentally measured spectrum.

To verify this assumption, a series of simulations was performed, which assume arbitrary $\tilde{m}_\text{InAs}^*$ value for both the single-particle \textit{nextnano++} simulation and for the hybrid model. The focus on the effective mass of the well material only was done (i) due to the fact that the conduction band eigenfunctions penetrate into the barrier only minimally and (ii) to reduce the number of free parameters (i.e. to minimize the chance of over-fitting). The optimal $\tilde{m}_\text{InAs}^* = 0.11\,m_0$ value was found, which minimizes the maximum error between the central frequency of the both $8$~nm and $18$~nm main peaks as predicted by the model and the central frequencies measured by the ellipsometry spectroscopy for the both samples, respectively. The imaginary part of the dielectric function $\epsilon(\omega)$ obtained this way is shown in Fig.~\ref{fig:single-band-fit}\,(a) for the $8$~nm QW and in Fig.~\ref{fig:single-band-fit}\,(b) for the $18$~nm QW with the dashed green lines. The experimentally measured spectra are also shown in Fig.~\ref{fig:single-band-fit}, as the solid red lines, for comparison.

It can be noticed that, while the position of the central frequency of the both peaks can be obtained more or less correctly with the help of $\tilde{m}_\text{InAs}^* = 0.11\,m_0$ in the case of the both systems, still the character of the spectrum predicted by the model is qualitatively incorrect in both cases. In the case of the $8$~nm QW, Fig.~\ref{fig:single-band-fit}\,(a), the pronounced discrepancy between the model and the experiment is the presence of an additional separated peak centered at about $1122$\rcm. This second peak, while smaller, is of comparable scale to the main one at the $1905$\rcm and could not have been lost in the noise or baseline of the ellipsometry measurement.
In the case of the $18$~nm QW, Fig.~\ref{fig:single-band-fit}\,(b), apart from the biggest peak at $919$\rcm, there is a series of smaller peaks at lower frequencies of: $447$\rcm, $602$\rcm, $447$\rcm and  $812$\rcm. The total area under the smaller peaks is non negligible in comparison to the biggest peak, which hardly can then be interpreted as a MSP mode. Furthermore, the smaller peaks contribute to significantly asymmetric overall shape of the spectrum, with the "center of mass" visibly shifted to the left of the $919$\rcm peak. These properties are absent from the experimental results as well as from the spectrum obtained with the help of the \ekdp, the shape of which is shown in Fig.~\ref{fig:single-band-fit} as the dotted black lines for comparison. The general character of the imaginary $\epsilon(\omega)$ spectra shown in Fig.~\ref{fig:single-band-fit}\,(a) and (b) is not particular to the specific $\tilde{m}_\text{InAs}^* = 0.11\,m_0$ value, but it is similar for a range of comparable $\tilde{m}_\text{InAs}^*$ values, for which the predicted position of the main peaks is close to the one observed experimentally. To sum up, there is no $\tilde{m}_\text{InAs}^*$ value, which allows to obtain both: an approximately correct peak position of the biggest peak and a spectrum with only a single main MSP peak without any additional smaller yet still significant plasmon modes. It can be concluded that it is not possible to obtain a correct qualitative picture, corresponding to the strong coupling between the inter-subband plasmon modes, by using arbitrary effective mass $\tilde{m}_\text{InAs}^*$.

The effective medium approximation adopted in this work relies on the effective medium width $W_\text{eff}$ parameter for obtaining the effective dielectric function $\tilde{\epsilon}_{zz}$ from the susceptibility $\chi_{zz}$ according to Eq.~(\ref{eq:chi_to_epsi}). The natural choice of $W_\text{eff} = 2 W_\text{GaSb} + W_\text{InAs}$ was used in the modeling, corresponding to the effective medium width equal to the physical width of the QW nanosystem. It may be noticed however that, with all other things being equal, changing the $W_\text{eff}$ will shift the central frequency of the peaks in the $\Im\left(\tilde{\epsilon}_{zz}\right)$ spectrum. A~natural question arises, whether it may be possible to obtain a good fit to the experimentally obtained spectrum for some $W_\text{eff} \ne 2 W_\text{GaSb} + W_\text{InAs}$. In this context, a brief discussion how $W_\text{eff}$ affects the $\Im(\tilde{\epsilon}_{zz})$ is in order. In the $W_\text{eff} \rightarrow \infty$ limit, it comes straightforwardly from Eq.~(\ref{eq:chi_to_epsi}) that
\begin{equation}
    \Im(\tilde{\epsilon}_{zz}) = \frac{\Im(\chi_{zz})}{W_{\text{eff}}},
\end{equation}
that is, the $\Im(\tilde{\epsilon}_{zz})$ and the $\Im(\chi_{zz})$ spectra are exactly proportional, with the $W_{\text{eff}}$ as the scaling factor. For example, if the spectrum consists of a single MSP mode, which absorbed all the individual ISP peaks due to strong coupling between the individual oscillators, then this single peak will have the same central frequency and the same FWHM in both the $\Im(\tilde{\epsilon}_{zz})$ and the $\Im(\chi_{zz})$ spectra. Departing from the $W_\text{eff} \rightarrow \infty$ limit and moving into direction of decreasing $W_\text{eff}$ leads to an increasing redshift in the position of the MSP peak in the mentioned example. Initially, this redshift does not affect significantly the character of the MSP peak, but below a certain value $W_\text{eff}$, the coupling of the individual ISP peaks effectively diminishes and the MSP peak starts to decompose into a series of peaks. It should be noticed that this behavior can already be observed in the simulation spectra (green dashed lines) in Fig.~\ref{fig:single-band-fit}\,(b). Having this in mind, it is clear that further decrease of the effective length width below $W_\text{eff} = 2 W_\text{GaSb} + W_\text{InAs}$ cannot lead to single MSP mode as observed in the experiment. Conversely, it may be possible with increasing $W_\text{eff}$.

In order to investigate this possibility, the $W_{\text{eff}}\to\infty$ limit is taken into account. Since $\Im(\tilde{\epsilon}_{zz}) \sim \Im(\chi_{zz})$ in this case, it is sufficient to fit $m^*_\text{InAs} = 0.12\,m_0$ considering the position of the main peak of $\Im(\chi_{zz})$ to the experimentally obtained positions of $\Im(\tilde{\epsilon}_{zz})$. The corresponding results are shown in Fig.~\ref{fig:single-band-fit}\,(c) and (d) for the QW widths of $8$~nm and $18$~nm, respectively, for fitted $m^*_\text{InAs} = 0.12\,m_0$. In the case of the larger QW, it is indeed possible to obtain a single MSP peak with a central frequency very similar to the one observed experimentally, namely $986$\rcm, and the presence of the additional peaks visibly reduced, as shown in Fig.~\ref{fig:single-band-fit}\,(d). However, the character of the simulated spectrum in Fig.~\ref{fig:single-band-fit}\,(c), for the $8$~nm QW with $W_{\text{eff}}\to\infty$, remains qualitatively similar to the simulated one in Fig.~\ref{fig:single-band-fit}\,(a), retaining both peaks at $1052$\rcm and $1849$\rcm, and qualitatively different from the experimental one.

To sum up, it is reasonable to conclude that plasmon description in type-III (broken gap) systems cannot be, in general, based on a single-particle single-band Hamiltonian restricted to the conduction Bloch bands, with a corresponding simple scalar effective mass $m^*_\text{QW}$ -- even an arbitrary one.

\bibliography{references.bib}
\end{document}